\documentclass[aps,prd,twocolumn,superscriptaddress,groupedaddress,nofootinbib]{revtex4}

\usepackage{amsmath,amssymb,bm,color,dcolumn,mathrsfs,tabularx,graphicx}
\usepackage[hidelinks=true]{hyperref}
\hypersetup{colorlinks=true,linkcolor=blue,citecolor=blue,urlcolor=blue}
\usepackage{Macro} 

\def\T{\Theta}
\def\dB{\hB^{\rm d}}
\def\ox{\ol{x}}
\def\Lg{\mC{L}_{\rm g}}
\def\o{\star}
\def\mg{\ave{\estg}}
\def\Re#1{{\rm Re}[#1]}
\def\uk{\ol{k}}

\begin{document}

\title{CMB internal delensing with general optimal estimator for higher-order correlations}

\author{Toshiya Namikawa}
\affiliation{Department of Physics, Stanford University, Stanford, California 94305, USA}
\affiliation{Kavli Institute for Particle Astrophysics and Cosmology, SLAC National Accelerator Laboratory, 2575 Sand Hill Road, Menlo Park, California 94025, USA}

\date{\today}

\begin{abstract}
We present a new method for delensing $B$ modes of the cosmic microwave background (CMB) using 
a lensing potential reconstructed from the same realization of the CMB polarization (CMB internal delensing).
The $B$-mode delensing is required to improve sensitivity to primary $B$ modes generated by, e.g., 
the inflationary gravitational waves, axion-like particles, modified gravity, primordial magnetic fields, 
and topological defects such as cosmic strings. However, the CMB internal delensing suffers from 
substantial biases due to correlations between observed CMB maps to be delensed and that used for 
reconstructing a lensing potential. 
Since the bias depends on realizations, we construct a realization-dependent (RD) estimator for correcting 
these biases by deriving a general optimal estimator for higher-order correlations. 
The RD method is less sensitive to simulation uncertainties. Compared to the previous $\l$-splitting method, 
we find that the RD method corrects the biases without substantial degradation of the delensing efficiency. 
\end{abstract}

\maketitle


\section{Introduction} \label{intro}

The cosmic microwave background (CMB) anisotropies are one of the best probe in cosmology. 
$B$ modes of the CMB polarization could be generated by the inflationary gravitational waves, 
axion-like particles, primordial magnetic fields, modified gravity, and 
topological defects such as cosmic strings 
(e.g., \Refs{Polnarev:1985,Lue:1999,Kamionkowski:1996zd,Seljak:1996gy,Seljak:2006,Shaw:2010,Pogosian:2011,Teng:2011xc}. 
In addition, the gravitational lensing converts some of the $E$ modes into $B$ modes \cite{Zaldarriaga:1998ar}, 
and this effect has been measured by multiple CMB experiments 
(e.g., \Refs{Hanson:2013daa,PB14:BB,ACT16:phi,P16:lensB,BKVIII}). 
The recent precise measurement of the CMB polarization by the BICEP2/Keck Array experiment indicates that 
the observed $B$ modes are dominated by the gravitational lensing $B$ modes \cite{BKVIII}. 
Thus, removal of the lensing-induced $B$ modes, usually referred to as {\it delensing}, 
will be soon required to improve sensitivity to the primary $B$ modes \cite{Kesden:2002ku,Knox:2002pe}.
The temperature and $E$-mode delensing are also important to constrain, e.g., 
the effective number of the relativistic species \cite{Green:2016}. 

Several works have discussed methods to remove the lensing contributions in the CMB anisotropies. 
\Refs{Seljak:2003pn,Smith:2010gu} provide a method to construct a template of the lensed CMB fluctuations 
and to remove it from the observed CMB fluctuations. 
\Ref{Green:2016} proposes a delensing technique to remove the lensing-induced CMB anisotropies 
including the higher-order terms of the lensing potential. 
\Refs{Simard:2015,Sherwin:2015} show the delensing method using the cosmic infrared background (CIB) 
(and this technique is demonstrated by \Refs{Larsen:2016,Manzotti:2017}).

The delensing methods, however, suffer from biases if the CMB maps are delensed with 
a lensing potential derived from the CMB maps (CMB internal delensing). 
In the CMB internal delensing, we use the same realization of the observed CMB anisotropies for 
the delensing and reconstruction of the lensing potential. 
Therefore, the CMB anisotropies to be delensed correlates with that used in 
the lensing potential reconstruction. 
As we discuss latter in \sec{bias}, this correlation substantially biases the power and scatter of 
the delensed CMB spectrum (delensing bias).

Methods to correct the delensing bias have been explored in several works. 
The delensing bias is pointed out by \Ref{Teng:2011xc} and is characterized by \Ref{Namikawa:2014a} 
(see also \Ref{Sehgal:2016} for the CMB temperature and \Ref{Carron:2017} for the Planck cases). 
\Refs{Teng:2011xc,Sehgal:2016} proposed a method to mitigate the delensing bias by 
splitting CMB anisotropies into annuli in multipole space so that CMB multipoles to be delensed and 
that used for the lensing reconstruction are uncorrelated. 
In this method, however, the efficiency of the delensing is degraded 
because we lose the CMB multipoles in the lensing reconstruction and 
precision of the reconstructed lensing potential decreases. 

In this paper, we study an alternative way of correcting the delensing bias 
to avoid degradation of the delensing efficiency and to estimate the delensed spectrum more accurately 
than using the simulation alone. To construct an estimator for the delensing bias, 
we use the similar analogy of the ``realization-dependent'' (RD) method in the lensing reconstruction 
\cite{Namikawa:2012,P13:phi,P15:phi}. 
In the reconstruction of the lensing potential, the leading-order bias in estimating 
the lensing-potential power spectrum comes from the disconnected part of the observed CMB trispectrum. 
In the RD method, the bias is estimated by combining both simulated and observed CMB fluctuations 
so that the estimator is optimal and less sensitive to the error in the simulated CMB covariance. 
The RD method in the lensing reconstruction is motivated by the optimal trispectrum estimator \cite{Regan:2010cn}. 

Compared to the lensing reconstruction, the delensing bias contains higher order correlations 
such as six point correlations. 
Thus, we first generalize the optimal trispectrum estimator to the case with high-order correlations (polyspectra). 
Then we employ the optimal polyspectra estimator to construct RD estimators for the delensing bias. 
We demonstrate that the RD method removes the delensing bias without degradation of the delensing efficiency. 
In this paper, for simplicity, we derive the RD method in the flat-sky approximation 
but the extension to the full sky approach is straightforward. 

This paper is organized as follows. 
\sec{bias} reviews the $B$-mode delensing and the delensing bias. 
\sec{opt} derives an optimal estimator for polyspectra based on the Edgeworth expansion of the CMB likelihood. 
\sec{RDdelens} provides the RD estimators for the delensing bias and shows results of numerical simulation. 
\sec{summary} is devoted to summary and discussion. 

\section{CMB internal delensing} \label{bias}

Here we briefly summarize notations in this paper and methods for the lensing reconstruction and delensing. 

\subsection{CMB polarization}

The lensing effect on the CMB anisotropies is expressed by a remapping of the primary CMB anisotropies. 
Denoting the primary CMB fluctuations at position $\hatn$ on the last scattering surface as $X(\hatn)$, 
where $X$ is either the CMB temperature $\T$, or polarization anisotropies $Q\pm\iu U$, 
the lensed CMB anisotropies in a direction $\hatn$, are given by (e.g., \Ref{Lewis:2006fu})
\al{
	\tX(\hatn) = X(\hatn+\bn\grad(\hatn)) \,. \label{eq:remap}
}
Here the two-dimensional vector, $\bn\grad(\hatn)$, is the deflection angle. 
The Fourier modes of the temperature anisotropies, and the $E$ and $B$ modes are defined as
\al{
	\T_{\bl} &= \FT{\hatn}{\bl}{f} \T(\hatn) \,, \\
	E_{\bl}\pm\iu B_{\bl} &= -\FT{\hatn}{\bl}{f} [Q\pm\iu U](\hatn) \E^{\mp 2\iu\varphi_{\bl}} \,, 
}
where $\varphi_{\bl}$ is the angle of $\bl$ measured from the $x$-axis. 
Substituting \eq{eq:remap} into the above equation gives the following expression of the lensed 
CMB anisotropies up to first order of the lensing potential \cite{Hu:2001kj}:
\al{
	\tT_{\bl} &= \T_{\bl} - \intl{\bL}\bL\cdot(\bl-\bL) \grad_{\bL}\T_{\bl-\bL}
	\,, \\
	\tE_{\bl} &= E_{\bl} - \intl{\bL}\grad_{\bL}(E_{\bl-\bL}z_{\bl\bL}-B_{\bl-\bL}w_{\bl\bL}) 
	\,, \\ 
	\tB_{\bl} &= B_{\bl} - \intl{\bL}\grad_{\bL}(B_{\bl-\bL}z_{\bl\bL}+E_{\bl-\bL}w_{\bl\bL}) 
	\,, \label{eq:lensed-EB}
}
where the weight functions are given by
\al{
	z_{\bl\bL} &= \bL\cdot(\bl-\bL) \cos 2(\varphi_{\bl-\bL}-\varphi_{\bl}) \,, \\
	w_{\bl\bL} &= \bL\cdot(\bl-\bL) \sin 2(\varphi_{\bl-\bL}-\varphi_{\bl}) \,.
}
If the primordial $B$ modes are ignored, the $B$ modes induced by gravitational lensing are expressed 
as a convolution of the lensing potential and primary $E$ modes. 

\subsection{Lensing reconstruction}

Precision of the lensing reconstruction in the future will be almost determined by 
the $EB$ quadratic estimator, and we focus on the case using the $EB$ estimator. 
For a fixed realization of the lensing potential in the universe, the lensing effect on CMB maps produces 
off-diagonal elements of the CMB covariance. From \eq{eq:lensed-EB}, 
the off-diagonal elements become \cite{Hu:2001kj}
\al{
	\ave{\tE_{\bl}\tB_{\bL-\bl}}_{\rm CMB} = -\tCEE_\l w_{\bL-\bl,\bL} \grad_{\bL} \,,
}
where $\ave{\cdots}_{\rm CMB}$ denotes the ensemble average over unlensed CMB anisotropies
with a fixed realization of the lensing potential $\grad$. 
We ignore the higher-order terms of the lensing potential and contributions from primary $B$ modes 
\footnote{
Note that the presence of the primordial gravitational-wave (GW) B modes does not affect 
the precision of the lensing reconstruction and delensing efficiency. 
This is because the GW B modes are significant only at the large scale ($\l<100$) and 
the lensing potential is reconstructed from small scale B modes.
}.
$\tCEE_\l$ is the lensed $E$-mode spectrum to mitigate the higher-order terms of 
the lensing-potential power spectrum \citep{Hanson:2010rp,Lewis:2011fk}.
The lensing estimators are then described as \cite{Hu:2001kj} 
\al{
	\estg^{EB}_{\bL} = A_L \Int{2}{\bl}{(2\pi)^2} 
		\frac{-\tCEE_\l w_{\bL-\bl,\bL}}{\hCEE_\l\hCBB_{|\bL-\bl|}} \hE_{\bl}\hB_{\bL-\bl} 
	\,. \label{Eq:quad}
}
Here $\hE$ ($\hB$) is the observed $E$ ($B$) modes, $\hCEE$ ($\hCBB$) is the observed $E$ ($B$) mode 
spectrum, and $A_L$ is the quadratic estimator normalization given by \Ref{Hu:2001kj}. 
In practical analysis, the filtering functions, $1/\hCEE_\l$ and $1/\hCBB_\l$, are generalized to the inverse of 
the covariance matrix of $E$ and $B$ modes, and are determined by simulation (e.g., see \Ref{Story:2014hni}). 
Unless otherwise stated, $\estg$ is equal to $\estg^{EB}$ with the correction of the mean-field bias, $\ave{\estg}$. 

\subsection{Delensing}

A method to remove lensing $B$ modes from observed $B$ modes is to make a template of the lensing $B$ modes
with a measured lensing potential and $E$ modes. Hereafter, we refer to this technique as the template method. 
The template of the lensing $B$ modes up to first order of the lensing potential is defined as \cite{Smith:2010gu}
\al{
	(\hE\o\estg)_{\bl} = -\Int{2}{\bl'}{(2\pi)^2} 
		\frac{\CEE_{\l'}}{\hCEE_{\l'}}\frac{\Cgg_{|\bl-\bl'|}}{\hCgg_{|\bl-\bl'|}} 
		w_{\bl,\bl-\bl'} \hE_{\bl'} \estg_{\bl-\bl'}
	\,. \label{Eq:Ephi}
}
Here the optimal weight is given as a product of the Wiener filters of the $E$ modes and lensing potential.
In practical analysis, the denominator of the Wiener filter is determined by simulations. 
This expression is derived by assuming that the reconstructed lensing potential is uncorrelated with 
the CMB anisotropies \cite{Smith:2010gu}. 
While it would be possible to make a template including this correlation, 
\eq{Eq:Ephi} is numerically tractable and useful for delensing analysis in practical cases. 
A lensing $B$-mode template to remove the higher-order terms of $\grad$ is also possible to construct 
but \eq{Eq:Ephi} is enough to reproduce most of the lensing $B$ modes \cite{Smith:2010gu}. 
The delensed $B$-mode power spectrum is then given by \cite{Smith:2010gu}
\al{
	C^{\dB\dB}_\l = \Int{2}{\bl'}{(2\pi)^2} w^2_{\bl,\bl-\bl'} 
		\left( 1-\frac{(\CEE_{\l'})^2}{\hCEE_{\l'}}\frac{(\Cgg_{|\bl-\bl'|})^2}{\hCgg_{|\bl-\bl'|}} \right) 
	\,. \label{Eq:resBB}
}
Alternatively, in high sensitivity polarization experiments, the delensed $B$ modes are obtained by 
the inverse remapping of the observed CMB map using the Wiener-filtered lensing potential \cite{Green:2016}. 
Hereafter, we call this approach the remapping method. 
While the remapping method is a simple way to remove higher order terms of the lensing potential, 
the inverse-remapping procedure simultaneously remaps noise fluctuations which should be taken into account 
for a noisy experiment \cite{Carron:2017}. 

\section{Delensing bias}

To delens $B$ modes, the observed $B$ modes to be delensed and that used in the lensing reconstruction 
are correlated and the delensed $B$ modes have significant biases. 
Here we discuss the bias in the delensed $B$-mode power spectrum. 

\subsection{CMB internal delensing with simulation}

\begin{figure}[t]
\bc
\includegraphics[width=8.5cm,height=6cm,clip]{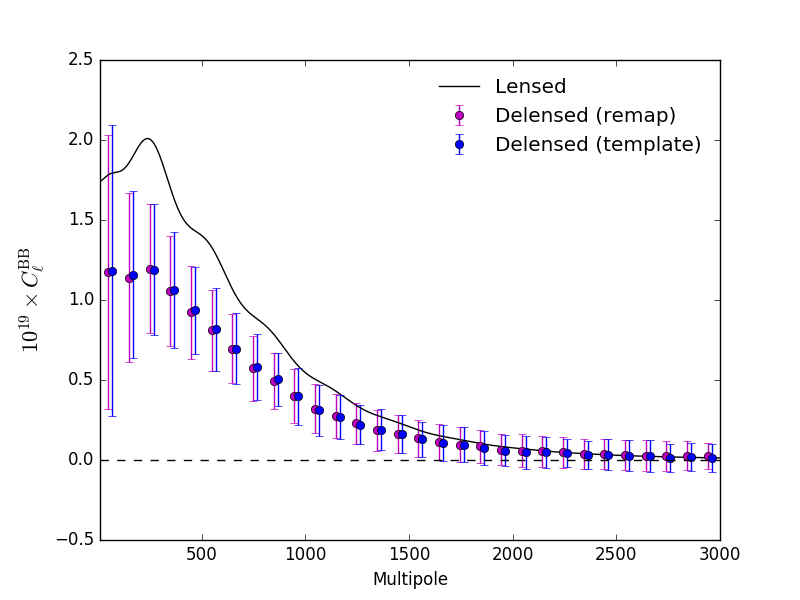} 
\caption{
Delensed $B$-mode power spectrum with the template and remapping methods. 
The solid line shows the lensing $B$-mode spectrum. 
The error bar denotes $1\sigma$ statistical error of one realization of observation in a $100$deg$^2$ patch. 
}
\label{fig:method}
\ec
\end{figure}

To study the CMB internal delensing, we simulate the lensed CMB, noise and reconstructed lensing potential maps 
as follows. 
We first generate $1200$ realizations of unlensed CMB and lensing potential maps in $10$deg$\times 10$deg region
as random Gaussian fields using CAMB output spectra \cite{Lewis:1999bs} 
\footnote{
The first $300$ realizations are used as ``observed" CMB fluctuations while the other $3\times 300$ realizations are 
used for estimating the (RD) delensing biases. 
}.
We assume the cosmological parameters consistent with \Ref{P15:main} without the inflationary B modes. 
The lensed CMB maps are obtained by remapping the unlensed CMB maps with the lensing potential \cite{Louis:2013}. 
The instrumental noise spectrum is computed with the formula of \Ref{Knox:1999} 
assuming $\sqrt{2}\times 3\muKac$ white noise in polarization map and $1$ arcmin Gaussian beam, 
and the noise maps are generated with this noise spectrum. 
The lensing potential is reconstructed from the lensed+noise (observed) maps with the $EB$ quadratic estimator 
using CMB multipoles in the range between $200$ and $3000$. 
We do not use the multipoles at $\l<200$ since the large-scale B modes are dominated by the Galactic foregrounds 
and are also not important in the lensing potential reconstruction. 
To characterize the delensing bias, we also use the input lensing potential with 
a random Gaussian noise whose power corresponds to the reconstruction noise
\footnote{
The leading order reconstruction noise is expressed as the disconnected bias, but the higher-order biases 
are not negligible in the following discussion. 
Thus we denote the reconstruction noise spectrum as $C_L^{\estg\estg}-C_L^{\grad\grad}$. 
}. 

We then perform delensing with the above lensing potentials and observed CMB maps. 
In the template method, the lensing $B$ modes are estimated from \eq{Eq:Ephi} and are subtracted from 
the observed $B$ modes. In the remapping method, the observed $Q$/$U$ maps are remapped by 
the Wiener-filtered lensing potential and we obtain the delensed $B$ modes from the delensed $Q$/$U$ maps. 
We do not apply Wiener filter for the CMB polarizations since its impact is negligible in our case. 

\subsection{Comparison between template and remapping methods}

\fig{fig:method} shows delensed $B$ modes with the template and remapping methods. 
For simplicity, we use the input lensing potential with the random Gaussian reconstruction noise 
since it does not create the delensing bias. 
The delensed $B$-mode power spectra using the template and remapping methods are similar 
since the lensing $B$ modes are approximately expressed as \eq{eq:lensed-EB} \cite{Challinor:2005jy}. 
In the following sections, we focus on the method of the (linear order) template method. 

\subsection{Delensing bias in $B$ mode power spectrum}

\begin{figure}[t]
\bc
\includegraphics[width=8.5cm,height=6cm,clip]{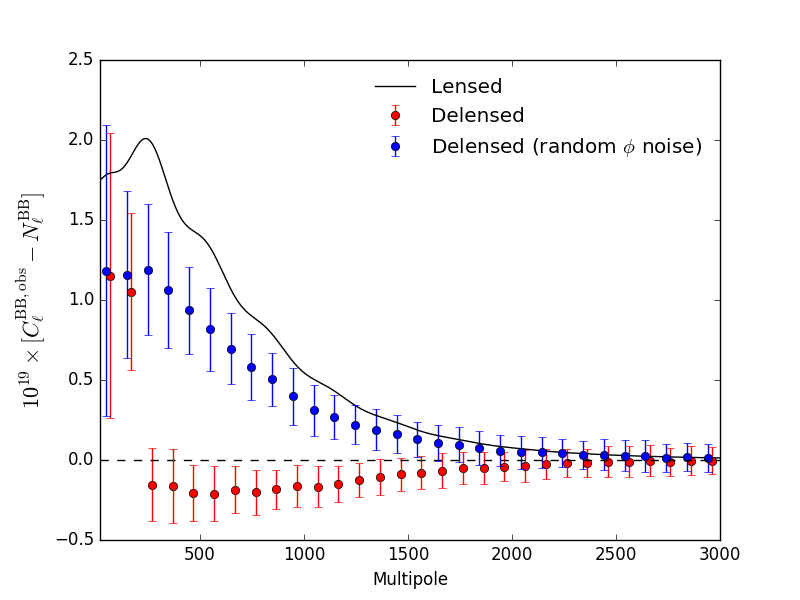} 
\caption{
Delensed $B$-mode power spectra using the lensing potential reconstructed from the CMB maps 
or the input lensing potential with the random Gaussian reconstruction noise.
The instrumental noise spectrum is subtracted from the delensed spectra. 
The blue points are equivalent to the $B$-mode spectrum with the template method shown in \fig{fig:method}. 
The error bars denote the $1\sigma$ statistical error for one realization of observation at a $100$deg$^2$ patch. 
}
\label{fig:del}
\ec
\end{figure}

\begin{figure*}[t]
\bc
\includegraphics[width=8.5cm,height=6cm,clip]{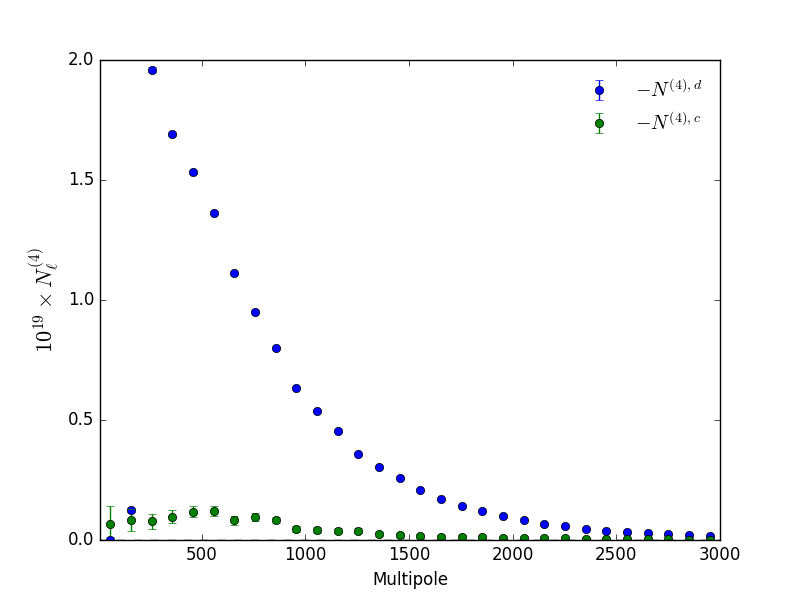} 
\includegraphics[width=8.5cm,height=6cm,clip]{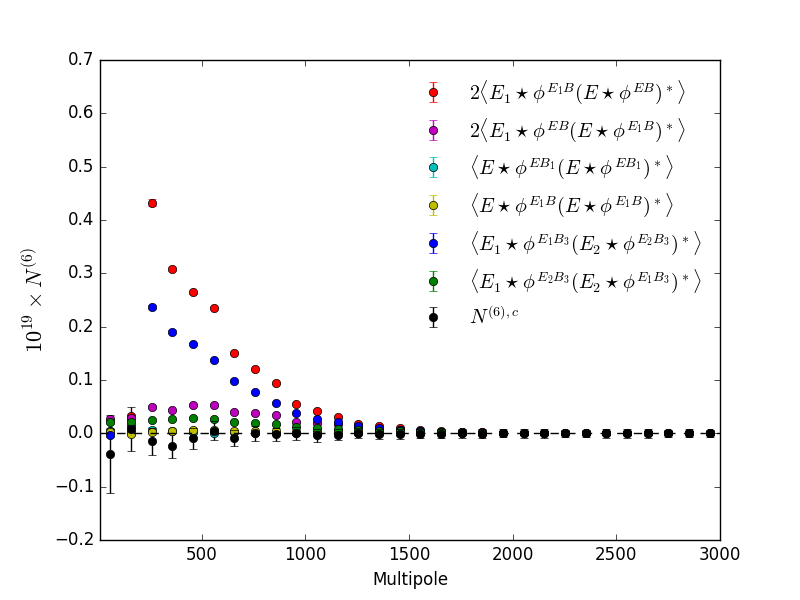} 
\caption{
Delensing biases from the four point (Left) and six point correlations (Right). 
The four point correlation is decomposed into the disconnected and connected parts. 
The six point correlation is divided into the seven terms described in \eq{Eq:N6}. 
The error bars denote the $1\sigma$ variance of the Monte Carlo simulation.  
}
\label{fig:contrib}
\ec
\end{figure*}

\begin{figure}[t]
\bc
\includegraphics[width=8.5cm,height=6cm,clip]{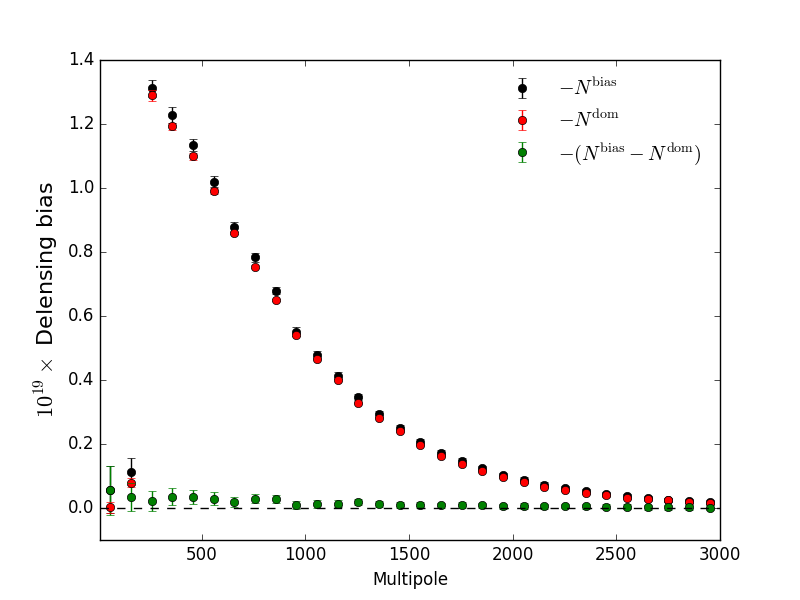} 
\caption{
Dominant contributions to the delensing bias defined as \eq{Eq:Ndom}. 
}
\label{fig:Ndom}
\ec
\end{figure}

Next we use the reconstructed lensing potential for delensing. 
\fig{fig:del} shows the delensed $B$-mode spectrum using the input lensing potential with the random 
Gaussian reconstruction noise or the reconstructed lensing potential. 
\footnote{
The instrumental noise power spectrum is subtracted from the delensed $B$-mode spectrum and thus 
the debiased spectra can be negative. 
}
Since the delensed $B$-mode spectrum with the random Gaussian reconstruction noise does not contain 
the delensing bias, the difference of the $B$-mode spectra between the two cases describes the delensing bias.
The discrepancy appears at $L\agt 200$ and the scatter of the power spectrum is also reduced. 

To understand the delensing bias, we expand the delensed $B$-mode spectrum into the four and six point correlations. 
If the lensing potential is estimated from an external data or the reconstruction noise is uncorrelated 
with the CMB anisotropies, the delensed $B$-mode power spectrum is simply equivalent to the power 
spectrum given in \eq{Eq:resBB}: 
\al{
	\ave{|\dB|^2} = \ave{|\hB-(\hE\o\estg')|^2} = (2\pi)^2\delta^{\rm D}_{\bm 0}C^{\dB\dB} \,. 
}
Here $\estg'$ is given as a sum of the true lensing potential and 
the random Gaussian reconstruction noise, $\delta^{\rm D}_{\bm 0}$ is the Delta function, and we omit the multipole $\bl$. 
In the CMB internal delensing, the delensed spectrum has additional contributions. 
The additional terms are given as 
\al{
	N^{\rm bias}
		&= \ave{|\hB-(\hE\o\estg)|^2} - \ave{|\hB-(\hE\o\estg')|^2} 
	\notag \\
		&= - 2\ave{\hB(\hE\o(\estg-\estg'))^*} 
	\notag \\
		&\quad + \ave{|(\hE\o\estg)|^2-|(\hE\o\estg')|^2}
	\notag \\
		&= N^{(4)} + N^{(6)}
	\,, \label{Eq:Nbias}
}
where we define
\al{
	N^{(4)} &\equiv - 2\ave{\hB(\hE\o(\estg-\estg'))^*} 
	\,, \\
	N^{(6)} &\equiv \ave{|(\hE\o\estg)|^2-|(\hE\o\estg')|^2} 
	\,, \label{Eq:N4N6}
}

The first term, $N^{(4)}$, is a four point correlation and leads to the following bias terms,
\al{
	N^{(4)} = -2\ave{\hB^*(\hE\o\estg)}_{\rm d}-2\ave{\hB^*(\hE\o(\estg-\grad))}_{\rm c}
	\,,
}
where $\ave{\cdots}_{\rm d}$ and $\ave{\cdots}_{\rm c}$ are the disconnected and connected parts, respectively. 

The second term, $N^{(6)}$, is a six point correlation and is decomposed into a connected term, 
products of the power spectrum and trispectrum, and products of three power spectra. 
To see this, from \eqs{Eq:Ephi,Eq:quad}, we explicitly rewrite the lensing $B$-mode template as
\al{
	(\hE\o\estg)_{\bl} &= \Int{2}{\bl'}{(2\pi)^2} \Int{2}{\bl''}{(2\pi)^2} 
		W_{\bl\bl'\bl''} \hE_{\bl'} \hE_{\bl''}\hB_{\bl-\bl'-\bl''} 
	\,,
}
where we define
\al{
	W_{\bl\bl'\bl''} \equiv \frac{\CEE_{\l'}\Cgg_{|\bl-\bl'|} w_{\bl,\bl-\bl'}}{\hCEE_{\l'}\hCgg_{|\bl-\bl'|}}
	\frac{A_{|\bl-\bl'|} \tCEE_{\l''} w_{\bl-\bl'-\bl'',\bl-\bl'}}{\hCEE_{\l''}\hCBB_{|\bl-\bl'-\bl''|}} 
	\,.
}
The power spectrum of the above quantity is the six point correlation, and we obtain
\al{
	&\ave{|(\hE\o\estg)|^2_{\bl}} = 
		\Int{2}{\bl'_1}{(2\pi)^2} \Int{2}{\bl''_1}{(2\pi)^2} 
		\Int{2}{\bl'_2}{(2\pi)^2} \Int{2}{\bl''_2}{(2\pi)^2} 
	\notag \\
		&\quad\times W_{\bl\bl'_1\bl''_1} W_{\bl\bl'_2\bl''_2} 
		\ave{\hE_{\bl'_1}\hE_{\bl''_1}\hB_{\bl-\bl'_1-\bl''_1}
		\hE^*_{\bl'_2}\hE^*_{\bl''_2}\hB^*_{\bl-\bl'_2-\bl''_2}}
	\,. \label{Eq:6pt-preexpand}
}
The disconnected part of the six point correlation is expanded as
\al{
	&\ave{\hE_{\bl'_1}\hE_{\bl''_1}\hB_{\bl-\bl'_1-\bl''_1}\hE^*_{\bl'_2}\hE^*_{\bl''_2}\hB^*_{\bl-\bl'_2-\bl''_2}}_{\rm d}
	\notag \\
	&= 2\ave{\hE_{\bl'_1}\hE_{\bl''_1}}\ave{\hB_{\bl-\bl'_1-\bl''_1}\hE^*_{\bl'_2}\hE^*_{\bl''_2}\hB^*_{\bl-\bl'_2-\bl''_2}}_{\rm c}
	\notag \\
	&+ 2\ave{\hE_{\bl'_1}\hE^*_{\bl''_2}}\ave{\hB_{\bl-\bl'_1-\bl''_1}\hE_{\bl''_1}\hE^*_{\bl'_2}\hB^*_{\bl-\bl'_2-\bl''_2}}_{\rm c}
	\notag \\
	&+ \ave{\hB_{\bl-\bl'_1-\bl''_1}\hB^*_{\bl-\bl'_2-\bl''_2}}\ave{\hE_{\bl'_1}\hE^*_{\bl'_1}\hE^*_{\bl'_2}\hE^*_{\bl''_2}}_{\rm c}
	\notag \\
	&+ \ave{\hE_{\bl'_1}\hE^*_{\bl''_2}}\ave{\hB_{\bl-\bl'_1-\bl''_1}\hE_{\bl''_1}\hE^*_{\bl'_2}\hB^*_{\bl-\bl'_2-\bl''_2}}_{\rm c}
	\notag \\
	&+ \ave{\hE_{\bl'_1}\hE_{\bl''_1}}\ave{\hE^*_{\bl'_2}\hE^*_{\bl''_2}}\ave{\hB_{\bl-\bl'_1-\bl''_1}\hB^*_{\bl-\bl'_2-\bl''_2}}
	\notag \\
	&+ \ave{\hE_{\bl'_1}\hE^*_{\bl'_2}}\ave{\hE_{\bl''_1}\hE^*_{\bl''_2}}\ave{\hB_{\bl-\bl'_1-\bl''_1}\hB^*_{\bl-\bl'_2-\bl''_2}}
	\,, \label{Eq:6pt-expand}
}
where we omit the terms irrelevant to the delensing bias (i.e., the terms contained in $|(\hE\o\estg')|^2$) 
and use the fact that the correlation between $E$ and $B$ modes vanishes.

The evaluation of the above equation is simplified by introducing an operator, $\ave{\cdots}_{\bm i}$, 
which is the ensemble average over quantities in $i$th set of realizations. 
For example, denoting $E_i$ and $B_i$ as the $E$ and $B$ modes in $i$th set of simulation, we can rewrite
the last term of \eq{Eq:6pt-expand} as
\al{
	&\ave{\hE_{\bl'_1}\hE^*_{\bl'_2}}\ave{\hE_{\bl''_1}\hE^*_{\bl''_2}}\ave{\hB_{\bl-\bl'_1-\bl''_1}\hB^*_{\bl-\bl'_2-\bl''_2}}
	\notag \\
	&= \ave{\hE_{1,\bl'_1}\hE^*_{1,\bl'_2}}_{\bm 1}\ave{\hE_{2,\bl''_1}\hE^*_{2,\bl''_2}}_{\bm 2}\ave{\hB_{3,\bl-\bl'_1-\bl''_1}\hB^*_{3,\bl-\bl'_2-\bl''_2}}_{\bm 3}
	\notag \\
	&= \ave{\hE_{1,\bl'_1}\hE^*_{1,\bl'_2}\hE_{2,\bl''_1}\hE^*_{2,\bl''_2}\hB_{3,\bl-\bl'_1-\bl''_1}\hB^*_{3,\bl-\bl'_2-\bl''_2}}_{\bm 1,2,3}
	\,,
}
where the index ($i=1,2,3$) denotes independent set of realizations. 
Substituting the above term into \eq{Eq:6pt-preexpand}, we obtain 
\al{
	\ave{|(\hE\o\estg)|^2} \supset \ave{(E_1\o\estg^{E_2B_3})(E_2\o\estg^{E_1B_3})^*}_{\bm 1,2,3}
	\,.
}
Here we again omit the multipole. 
Substituting \eq{Eq:6pt-expand} into \eq{Eq:6pt-preexpand}, and 
including the connected part of the six point correlation, we obtain
\al{
	N^{(6)} 
		&= 2\ave{\ave{E_1\o\estg^{E_1B_2}(E_2\o\estg^{E_2B_2})^*}_{\bm 1}}_{2,\rm c}
		\notag \\
		&\quad + 2\ave{\ave{E_1\o\estg^{E_2B_2}(E_2\o\estg^{E_1B_2})^*}_{\bm 1}}_{2,\rm c}
		\notag \\
		&\quad + \ave{\ave{E_2\o\estg^{E_2B_1}(E_2\o\estg^{E_2B_1})^*}_{\bm 1}}_{2,\rm c}
		\notag \\
		&\quad + \ave{\ave{E_2\o\estg^{E_1B_2}(E_2\o\estg^{E_1B_2})^*}_{\bm 1}}_{2,\rm c}
		\notag \\
		&\quad + \ave{(E_1\o\estg^{E_1B_3})(E_2\o\estg^{E_2B_3})^*}_{\bm 1,2,3} 
		\notag \\
		&\quad + \ave{(E_1\o\estg^{E_2B_3})(E_2\o\estg^{E_1B_3})^*}_{\bm 1,2,3}
		\notag \\
		&\quad + \ave{|\hE\o\estg|^2}_{\rm c} 
	\,. \label{Eq:N6}
}
Here $\ave{\cdots}_{i,\rm c}$ extracts the connected part of the ensemble average. 
This expression is convenient to evaluate the bias using simulation.  
In the above equation, the first four terms are the product of 
the trispectrum and power spectrum, and the next two terms are the product of the three power spectra. 
The last term contains the pentaspectrum. 

\fig{fig:contrib} shows the power spectra of each contribution in the four point and six point delensing bias. 
The most significant contribution comes from the disconnected part of the four point correlation $N^{(4),\rm d}$. 
Some terms involved in $N^{(6)}$ are also significant. The last term in \eq{Eq:N6} is negligible. 

\fig{fig:Ndom} plots the dominant contribution of the delensing bias which is given by
\al{
	N^{\rm dom} &= -2\ave{\hB^*\hE\o\estg}_{\rm d}
		\notag \\
		&\quad + 2\ave{\ave{(\hE_1\o\estg^{E_1B_2})(\hE_2\o\estg^{E_2B_2})^*}_{\bm 1}}_{2,\rm c}
		\notag \\
		&\quad + \ave{(\hE_1\o\estg^{E_1B_3})(\hE_2\o\estg^{E_2B_3})^*}_{\bm 1,2,3}
	\,. \label{Eq:Ndom}
}
$N^{(4),\rm c}$ cancels with a term involved in $2\ave{\ave{E_1\o\estg^{E_2B_2}(E_2\o\estg^{E_1B_2})^*}_{\bm 1}}_{2,\rm c}$. 
The above expression can be derived by rewriting the delensed $B$ modes as \cite{Namikawa:2014a} 
\al{
	\dB \simeq \hB - (\hE\o\estg') - \ave{(\hE_1\o\estg^{E_1B})}_{\bm 1} \,. \label{Eq:dB}
}
Because the third term is proportional to $\hB$ \cite{Namikawa:2014a}, 
the bias depends on the realization of the observed $B$ mode. 
The primary $B$ modes which we want to detect are also simultaneously subtracted.

\subsection{Mitigating delensing bias}

Let us discuss how to mitigate the delensing bias. 

\begin{figure*}[t]
\bc
\includegraphics[width=8.5cm,height=6cm,clip]{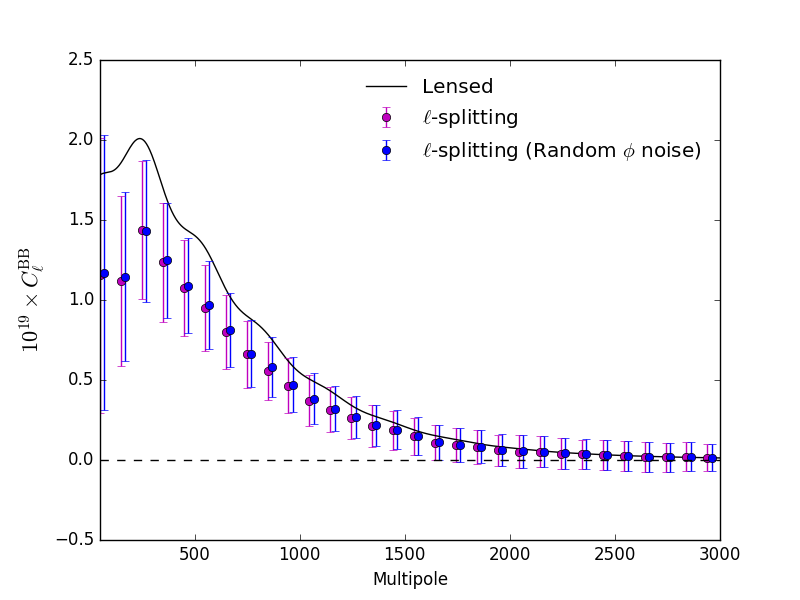} 
\includegraphics[width=8.5cm,height=6cm,clip]{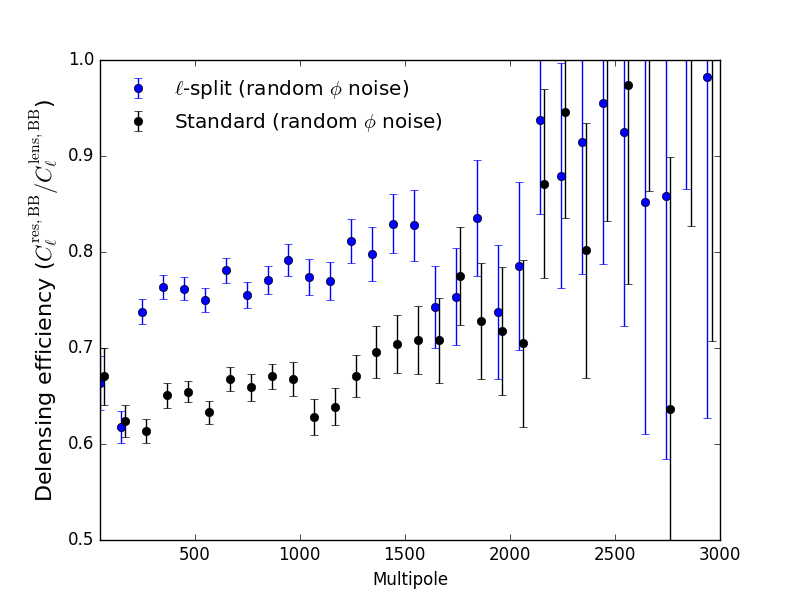} 
\caption{
{\it Left}: 
Same as Fig.~\ref{fig:del} but for the $\l$-splitting method. 
{\it Right}:
Delensing efficiency of the $B$-mode polarization. 
We show the standard and $\l$-splitting methods with the random Gaussian reconstruction noise. 
The error bar denotes the $1\sigma$ simulation error. 
}
\label{fig:lsplit}
\ec
\end{figure*}

Several works have discussed a method to avoid the delensing bias by splitting multipole range 
of CMB into annuli in $\l$ space so that CMB maps to be delensed and that used for 
the lensing reconstruction are uncorrelated \cite{Teng:2011xc,Sehgal:2016}. 
Fig.~\ref{fig:lsplit} shows the delensed $B$-mode spectra using the $\l$-splitting method 
of \Ref{Teng:2011xc} where CMB multipole is split into four annuli with equal spacing between $200$ and $3000$. 
We also show the delensing efficiency using the standard and the $\l$-splitting methods 
where we use the random Gaussian reconstruction noise to avoid the delensing bias. 
The efficiency of delensing is defined as the ratio of the residual lensing $B$-mode spectrum to 
the original lensing-$B$ mode spectrum, indicating that how significantly the lensing $B$ mode is removed.
In the $\l$-splitting method, the delensing bias (difference between cyan and magenta points) 
is not significant. 
However, comparing the standard method (black) with the $\l$-splitting technique (blue) 
in the random $\grad$ noise cases, the use of the $\l$-splitting method decreases the delensing efficiency. 
This is because the CMB multipoles used for the lensing reconstruction decrease and 
the precision of the lensing potential reconstruction is degraded. 

As an alternative method, in the following sections, 
we discuss the RD method which utilizes observed CMB maps to estimate 
the delensed $B$-mode power spectrum without degradation of the delensing efficiency and 
is more reliable than using simulation alone. 
This method emerges naturally as the delensing bias depends on observed $B$ mode. 
To construct an estimator, we employ the similar analogy of the RD method 
for reconstructing the lensing potential \cite{Namikawa:2012} 
which is motivated by the optimal trispectrum estimator. 
We extend this technique to the delensing bias by deriving the optimal estimator 
for the polyspectra in \sec{opt}, and applying the optimal polyspectra estimator 
to the disconnected part of the delensing bias which has most significant contribution in the bias 
in \sec{RDdelens}. 

\section{Optimal estimator for polyspectra} \label{opt}

\subsection{Derivation}

We consider a set of variables $\bm{x}=(x_1,x_2,\dots,x_N)$. 
For example, the index of $x_i$ denotes the multipoles ($\bl$) in the flat sky analysis and 
the harmonic coefficients ($\l m$) in the full sky analysis, 
and specifies the temperature, $E$ or $B$ modes. 
Expanding the likelihood of these variables around the Gaussian likelihood, a term 
which contains $n$th order of the correlation is given by (e.g., \Refs{Amendola:1996,Regan:2010cn})
\al{
	\mC{L} \propto \sum_{i_j} k_{i_1\dots i_n} \PD{}{x_{i_1}}\cdots\PD{}{x_{i_n}}\Lg \,,
}
where $k_{i_1\dots i_n}$ is the coefficients of the $n$th order derivative, $i_1,\cdots,i_n\in [1,N]$, 
and the operator $\sum_{i_j}$ sums over all possible combinations of $i_1,\cdots,i_n$. 
For example, if $n=6$, $k$ is the sum of the six point correlation and products of the three point correlations. 
The Gaussian likelihood is given as
\al{
	\Lg \propto \exp\left(-\frac{1}{2}\sum_{ij}c_{ij}x_ix_j\right) \,, 
}
where $c_{ij}=(C^{-1})^{x_ix_j}$ with $C$ denoting the covariance between $\bm{x}$. 

To derive an explicit expression of the derivative of the Gaussian likelihood 
with respect to $x_{i_1},x_{i_2},\dots,x_{i_n}$, we first define the following quantities: 
\al{
	p^n_m = \ave{\ox^{\bm 1}_{i_1}\cdots\ox^{\bm m}_{i_{2m}}\ox_{i_{2m+1}}\cdots\ox_{i_n} 
		+ \text{(perms.)} }_{\bm{1,\cdots,m}}
	\,. \label{Eq:pnm}
}
Here $m$ is a non-negative integer and $p^n_m=0$ if $m>[n/2]$. 
Again, to simplify the numerical calculation, we introduce the operator, $\ave{\cdots}_{\bm{i}}$, 
which takes the ensemble average over $i$th set of realizations, $\ox^{\bm{i}}$, and 
$\ox$ is an inverse-variance filtered variable fields as
\al{
	\ox_j = \sum_i c_{ij}x_i \,.
}
For example, if $n=6$, we obtain
\al{
	p^6_0 &= \ox_{i_1}\cdots\ox_{i_6}
	\,, \\
	p^6_1 &= \ave{\ox^{\bm 1}_{i_1}\ox^{\bm 1}_{i_2}\ox_{i_3}\cdots\ox_{i_6} + \text{(14 perms.)}}_{\bm 1}
	\,, \label{Eq:p61} \\
	p^6_2 &= \ave{\ox^{\bm 1}_{i_1}\cdots\ox^{\bm 2}_{i_4}\ox_{i_5}\ox_{i_6} + \text{(44 perms.)}}_{\bm 1,2}
	\,, \label{Eq:p62} \\
	p^6_3 &= \ave{\ox^{\bm 1}_{i_1}\cdots\ox^{\bm 3}_{i_6} + \text{(14 perms.)}}_{\bm 1,2,3}
	\,, \label{Eq:p63}
}
and $p^6_m=0$ for $m>3$. Using
\al{
	\PD{\ox_j}{x_i} = c_{ij} = \ave{\ox_i\ox_j} \,,
}
the quantities $p^n_m$ satisfy
\al{
	p^{n+1}_0     &= \ox_{i_{n+1}} p^n_0 \,, \notag \\
	p^{n+1}_{m+1} &= \PD{p^n_m}{x_{i_{n+1}}} + \ox_{i_{n+1}} p^n_{m+1} \,. \label{Eq:pnm:rel}
}

With $p^n_m$, we find that the $n$th order derivative of the Gaussian likelihood is simply expressed as
\al{
	(-1)^n\PD{^n\Lg}{x_{i_1}\cdots\pd x_{i_n}} = \Lg \sum_{m=0} (-1)^m p^n_m
	\quad (n\geq 1) \,. \label{Eq:dlnL/dx}
}
To show this, we first consider the case with $n=1$. The left-hand side of the above equation becomes 
\al{
	-\PD{\Lg}{x_{i_1}} = \ox_{i_1} \Lg \,, \label{Eq:dlnL/dx:n1}
}
On the other hand, the quantity defined in \eq{Eq:pnm} is given by $p^1_0=\ox_{i_1}$ and 
the right-hand side of \eq{Eq:dlnL/dx} coincides with \eq{Eq:dlnL/dx:n1}. 
To show the case with $n$ greater than $1$, we differentiate \eq{Eq:dlnL/dx} with respect to $x_{i_{n+1}}$ and obtain
\al{
	&\frac{(-1)^n\pd^n\Lg}{\pd x_{i_1}\dots\pd x_{i_{n+1}}} 
		= \sum_{m=0} (-1)^m \left[ \Lg \PD{p^n_m}{x_{i_{n+1}}} + p^n_m \PD{\Lg}{x_{i_{n+1}}} \right]
	\notag \\ 
	&\qquad = \sum_{m=0} (-1)^m \Lg \PD{p^n_m}{x_{i_{n+1}}} - \sum_{m=0} (-1)^m \Lg p^n_m \ox_{i_{n+1}}
	\,.
}
Replacing $m$ with $m+1$ in the second term, we obtain
\al{
	\frac{(-1)^n\pd^n\Lg}{\pd x_{i_1}\dots\pd x_{i_{n+1}}} 
		&= \sum_{m=0} (-1)^m \Lg \left[ \PD{p^n_m}{x_{i_{n+1}}} + p^n_{m+1} \ox_{i_{n+1}} \right] 
	\notag \\ 
		&\quad - \Lg \ox_{i_{n+1}}p^n_0 
	\,. \label{Eq:opt:1}
}
Using \eq{Eq:pnm:rel}, the first term becomes
\al{
	\sum_{m=0} (-1)^m \Lg p^{n+1}_{m+1} = \sum_{m=0} (-1)^{m+1} \Lg p^{n+1}_m + \Lg p^{n+1}_0
	\,. \label{Eq:opt:2}
}
Substituting \eq{Eq:opt:2} into the first term of \eq{Eq:opt:1} leads to \eq{Eq:dlnL/dx} 
and \eq{Eq:dlnL/dx} is satisfied for any $n$. 

The optimal estimator of the $n$th order polyspectra, $k$, is proportional to the derivative 
of the likelihood with respect to $k$ as similar to the trispectrum \cite{Namikawa:2012}. 
From \eq{Eq:dlnL/dx}, the estimator of $k$ is given as
\al{
	\hk_{i_1\cdots i_n} 
		&= A_{i_1\cdots i_n}\left[p^n_0 + \sum_{m=1} (-1)^{m+n} p^n_m\right] 
		\label{Eq:opt:k} \\
		&\equiv A_{i_1\cdots i_n} \uk_{i_1\cdots i_n}
		\,, \label{Eq:ukdef}
}
where $A_{i_1\cdots i_n}$ is a normalization of the estimator so that the estimator is unbiased and 
$\uk_{i_1\cdots i_n}$ is the unnormalized estimator. 
The second term in the parenthesis is the estimator of the disconnected part of $k$:
\al{
	\hk^{\rm d}_{i_1\cdots i_n} = -A_{i_1\cdots i_n}\sum_{m=1} (-1)^{m+n} p^n_m
	\,. \label{Eq:opt:kd}
}
If the products of the polyspectra in $k$ are negligible 
(e.g., the product of the three-point correlations in $k^{i_1\cdots i_6}$), 
\eq{Eq:opt:k} becomes the estimator for the connected part of the $n$-point correlation. 

Note that an optimal estimator for the amplitude of the $n$-point correlation is defined as
\al{
	\widehat{A} \equiv \left[\sum_{i_j}\frac{k_{i_1\cdots i_n}^2}{\sigma^2_{i_1\cdots i_n}}\right]^{-1}
		\sum_{i_j}\frac{k_{i_1\cdots i_n}}{\sigma^2_{i_1\cdots i_n}} \hk_{i_1\cdots i_n}
	\,,
}
where the variance of the $n$-point correlation, $\sigma^2_{i_1\cdots i_n}$, 
is usually assumed to be the product of the two point-correlations. 

\subsection{Suboptimal estimator}

Here we discuss the accuracy of the estimator given in \eq{Eq:opt:k}. 
The estimator requires a set of simulated data, $\ox^{\bm i}$. 
In the trispectrum case ($n=4$), when we use an incorrect covariance 
$\ave{\ox_i\ox_j}=\ave{\ox_i\ox_j}^{\rm true}+\delta\ave{\ox_i\ox_j}$ in the simulation,
the estimator of \eq{Eq:opt:k} has no contribution at linear order of $\delta \ave{\ox_i\ox_j}$ \cite{Namikawa:2012}. 
Thus, the estimator is less sensitive to a mismatch between the simulated and observed covariance 
and more accurate than correcting the disconnected bias with the simulation alone. 

In general cases, however, the estimator has contributions from linear order of $\delta \ave{\ox_i\ox_j}$. 
For example, in $n=6$, the following terms involved in \eq{Eq:p61} produce biases:
\al{
	\ave{p^6_1} \ni \ave{\ox_{i_1}\ox_{i_2}}\ave{\ox_{i_3}\cdots\ox_{i_6}}_{\rm c} + \text{(14 perms.)}
	\,. \label{Eq:opt:p61:bias}
}
A possible way to mitigate these terms is to redefine $p^6_1$ as
\al{
	(p^6_1)' &\equiv p^6_1 
		- \ave{\ox_{i_1}\ox_{i_2}}\ave{\ox_{i_3}\cdots\ox_{i_6}}_{\rm c} + \text{(14 perms.)}
	\notag \\
		&\quad + \ox_{i_1}\ox_{i_2}\ave{\ox_{i_3}\cdots\ox_{i_6}}_{\rm c} + \text{(14 perms.)}
	\,. 
}
The second term is added so that its ensemble average cancels with \eq{Eq:opt:p61:bias}, while 
the third term is introduced such that the additional terms have zero mean. 
Combining $p^6_m$ in \eqs{Eq:p61,Eq:p62} and \eqref{Eq:p63}, 
the estimator for the disconnected part of the polyspectra is given by 
\al{
	&\hk^{\rm d,sub}_{i_1\cdots i_n} \propto (p^6_1)'-p^6_2+p^6_3 
		\notag \\
		&\quad = [(\ox_{i_1}\ox_{i_2}-\ave{\ox_{i_1}\ox_{i_2}})\ave{\ox_{i_3}\cdots\ox_{i_6}} + \text{(14 perms.)}] 
		\notag \\
		&\quad\qquad + p_1^6 - 2p^6_2 + 4p^6_3
	\,. \label{Eq:subopt}
}
Note that we discuss an alternative way to derive the above estimator in Appendix \ref{app:suboptderiv} 
by generalizing the discussion in \Ref{Carron:2017} to the polyspectra. 
The above estimator has no contributions from linear order of $\delta \ave{\ox_i\ox_j}$ and 
also from error of the trispectrum at linear order, $\delta\ave{\ox_i\ox_j\ox_k\ox_l}_{\rm c}$.
Thus, the suboptimal estimator of \eq{Eq:subopt} is more accurate than the estimator defined in \eq{Eq:opt:kd}. 

\section{Realization-dependent delensing} \label{RDdelens}

We here discuss the RD estimators for the removal of the delensing bias. 
Since the dominant contribution to the delensing bias comes from the disconnected part, 
we apply the above optimal polyspectra estimators given in \eqs{Eq:opt:kd,Eq:subopt} to 
the disconnected part of the delensing bias. 
Since the derivation requires a lengthy calculation, here we only show the results 
and the details of the calculation are given in Appendix \ref{app:bias}.

From \eqs{Eq:Nbias,Eq:N4N6}, the bias is described as
\al{
	N^{\rm bias} = N^{(4)} + N^{(6)} \,.
}
$N^{(4)}$ is a four point correlation and its disconnected part is the most dominant contribution 
in the delensing bias. The disconnected part of $N^{(4)}$ can be replaced with the following RD estimator 
(see Appendix \ref{app:4pt-bias}); 
\al{
	\hN^{(4),d} = -2\ave{[\hE_1\o\estg+\hE_{012}\o(\estg^{E_{0,-2}B_1}+\estg^{E_1B_{0,-2}})]\hB^*_{01}}_{\bm 1,2}
	\,. \label{Eq:hN4d}
}
Here we define $X_{i,\pm j\dots}=X_i\pm X_j+\dots$ and $i=0$ means the real data. 
$N^{(6)}$ is a six point correlation and its disconnected part is the dominant source of the bias. 
According to the optimal polyspectra estimator, the RD estimator for 
the disconnected part of the six point correlation is given by 
$\hN^{(6)}\propto p^6_1-p^6_2+p^6_3$ where $p^6_i$ is defined in \eq{Eq:pnm} with four $E$ and two $B$ modes. 
We find that the RD estimator for the disconnected part of the six point correlation becomes 
(see Appendix \ref{app:6pt-opt})
\al{
	\hN^{(6),d} &= \ave{\Re{\hE_{012}\o(2\estg_{01}+2\estg_{13})\mC{B}_{0123}^*}}_{\bm 1,2,3}
		\notag \\
		&\quad + \ave{\hE_{0,-3}\o\estg_{11}(\hE_{03}\o\estg_{11})^*}_{\bm 1,3}
	\,, \label{Eq:hN6d}
}
where we define $2\estg_{ij}=\estg^{E_iB_j}+\estg^{E_jB_i}$ and 
\al{
	\mC{B}_{0123} &\equiv \hE_{012}\o(2\estg-2\estg_{02}-4\estg_{03}+2\estg_{23})
		\notag \\
		&\quad -2\hE_3\o\estg+\hE_{0,-2}\o(2\estg_{01}-\estg_{13})
	\,. 
}

As discussed in the previous section, the suboptimal estimator of \eq{Eq:subopt} is more accurate 
than the estimator of \eq{Eq:hN6d}. The suboptimal RD estimator for the six point correlation is given by 
(see Appendix \ref{app:6pt-sub})
\al{
	\hN^{(6),\rm d,sub} = \ave{\hE_{012}\o(2\estg_{01}+2\estg_{13})(\mC{B}_{0123}^{\rm sub})^*}_{\bm 1,2,3}
	\,, \label{Eq:hN6d:sub}
}
where we define 
\al{
	\mC{B}^{\rm sub}_{0123} &\equiv 
		\hE_{012}\o(2\estg-4\estg_{02}-8\estg_{03}+8\estg_{23}+2\estg_{11})
	\notag \\
		&\quad -2\hE_3\o(2\estg+\estg_{11}+\estg_{33}) 
	\notag \\
		&\quad + 2(\hE_{01}-2\hE_2)\o(\estg_{01}-\estg_{13})
	\,.
}

As shown in \fig{fig:Ndom}, the dominant terms of the delensing bias are given by \eq{Eq:Ndom}. 
The RD estimator for \eq{Eq:Ndom} can be constructed by extracting the corresponding terms in 
the suboptimal estimator. The details are given in Appendix \ref{app:bias} and the result is
\al{
	&\hN^{(6),\rm d,dom} 
		= 2\hE_{0,-1}\o\estg^{E_{01}B_{02}}[\hE_{23}\o(\estg_{22}-2\estg_{23})]^*
	\notag \\
		&\quad +\hE_1\o\estg^{E_{01}B_{02}}[2\hE\o(\estg-\estg_{22}-\estg^{E_2B})
	\notag \\
		&\quad -\hE_2\o(\estg^{(2E+E_2)B_{02}}-\estg_{22})+2\hE_3\o\estg^{E_2B_3}]^*
	\,.
}

The above RD estimators are more reliable than the use of the delensing bias estimated from simulation alone. 
In the followings, we check how the use of $\hN^{(6),\rm d,sub}$ or $\hN^{(6),\rm d,dom}$ degrades 
the performance of the delensing. 

\begin{figure}[t]
\bc
\includegraphics[width=8.5cm,height=6cm,clip]{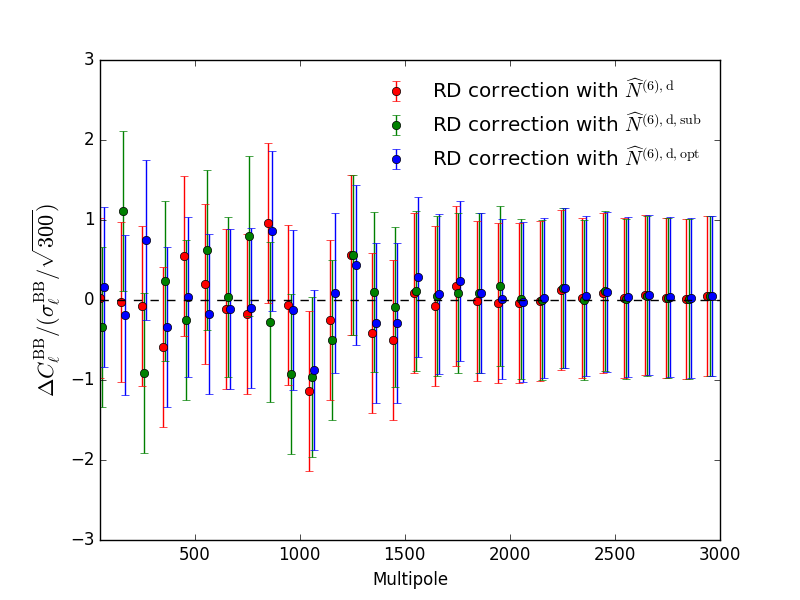} 
\caption{
Difference between the $B$-mode spectra using the RD estimators and with the random Gaussian reconstruction noise, 
divided by the $1\sigma$ simulation error.
We show the cases where the delensing bias is estimated using $\hN^{(6),\rm d}$, $\hN^{(6),\rm d,sub}$ 
or $\hN^{(6),\rm d,dom}$. 
}
\label{fig:rd}
\ec
\end{figure}

\begin{figure}[t]
\bc
\includegraphics[width=8.5cm,height=6cm,clip]{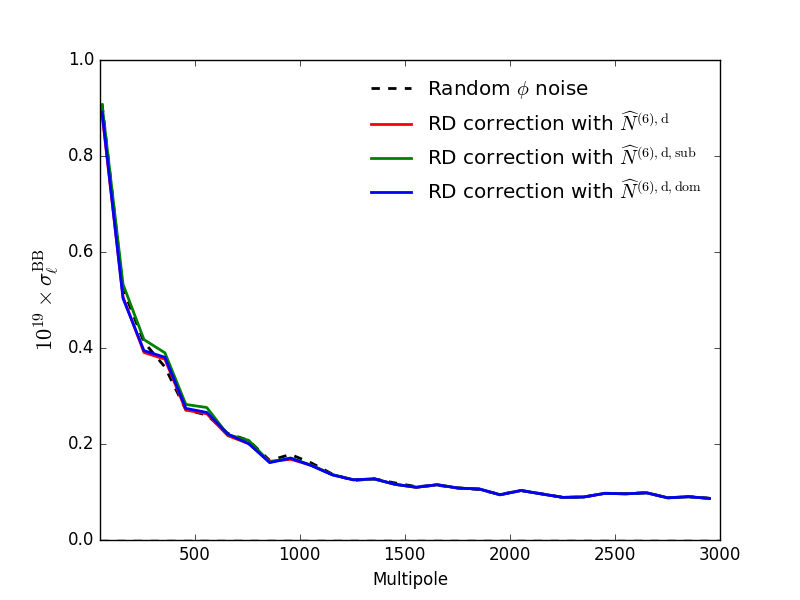} 
\caption{
$1\sigma$ statistical error of the delensed $B$-mode power spectrum with corrections of 
the delensing bias using the RD estimators. We also show the case with the random Gaussian reconstruction noise. 
}
\label{fig:rdvar}
\ec
\end{figure}

\fig{fig:rd} shows the difference between the $B$-mode power spectra using the RD estimators and with 
the random Gaussian reconstruction noise, divided by the $1\sigma$ simulation error. 
After the correction of the delensing bias with the RD estimators, 
the delensed spectrum becomes consistent with that in the case with the random Gaussian reconstruction noise.
\fig{fig:rdvar} shows the $1\sigma$ statistical error of the delensed $B$-mode spectrum after the bias correction. 
The statistical errors are almost unchanged between the three RD estimators. 
We conclude that the use of the suboptimal RD estimators ($\hN^{\rm sub}$ or $\hN^{\rm dom}$) does not degrade 
the performance of the delensing. 

\section{Summary and discussion} \label{summary}

We have derived the RD estimators to correct the delensing bias.
We first showed that the delensing bias is significant in the delensed $B$-mode power spectrum. 
To correct the bias, we derived the optimal polyspectra estimator in general cases. 
We then formulated the RD estimators for the delensing bias with the polyspectra estimator. 
Unlike the $\l$ splitting method proposed in previous works, 
the RD method corrects the delensing bias without substantial loss of the delensing efficiency. 

A similar bias on $B$ modes discussed in this paper would be appeared when we remove, for example,  
the spatially varying components of foregrounds estimated from the same realization of CMB maps with 
a quadratic estimator. 
The RD method presented in this paper would be also useful to mitigate such biases. 

We have focused on the $B$-mode delensing but the delensing bias is also significant in 
the CMB temperature and $E$-mode delensing \cite{Sehgal:2016,Carron:2017} (see also Appendix \ref{app:other}). 
Unlike the $B$-mode delensing focused in this paper, the higher-order terms of the lensing potential 
is not negligible in the temperature an $E$-mode delensing, and the remapping method or the template method 
including the higher-order terms of the lensing potential is more efficient than the linear-order template method. 
The delensing bias contains non-negligible contributions from the higher-order correlations 
(e.g. eight-point correlation of the observed CMB maps). 
To apply the RD method described in this paper to the temperature and $E$-mode delensing, 
we need to include the optimal polyspectra up to more than six point correlations. 

In this paper, we have focused on the application of the optimal polyspectra estimator to the delensing. 
The estimator is also useful to estimate, e.g., the bispectrum of the reconstructed 
lensing potential which is a six point correlation of the observed CMB anisotropies, and 
higher-order statistics of the kinetic SZ effects. 
The lensing bispectrum is generated by the nonlinear growth of the large-scale structure and 
the post-Born corrections. 
Since measurements of the lensing bispectrum will be useful to extract additional cosmological 
information in future CMB experiments \cite{Namikawa:2016b,Boehm:2016,Pratten:2016,Liu:2016nfs}, 
we will investigate the feasibility of the RD method for a lensing bispectrum measurement in our future work. 

\begin{acknowledgments}
We would like to thank Chao-Lin Kuo for support of this work and Antony Lewis for helpful comments. 
This research used resources of the National Energy Research Scientific Computing Center, 
which is supported by the Office of Science of the U.S. Department of Energy under Contract No. DE-AC02-05CH11231.
\end{acknowledgments}

\onecolumngrid
\appendix

\section{Delensing bias in other cases} \label{app:other}

\begin{figure*}[t]
\bc
\includegraphics[width=8.7cm,height=6cm,clip]{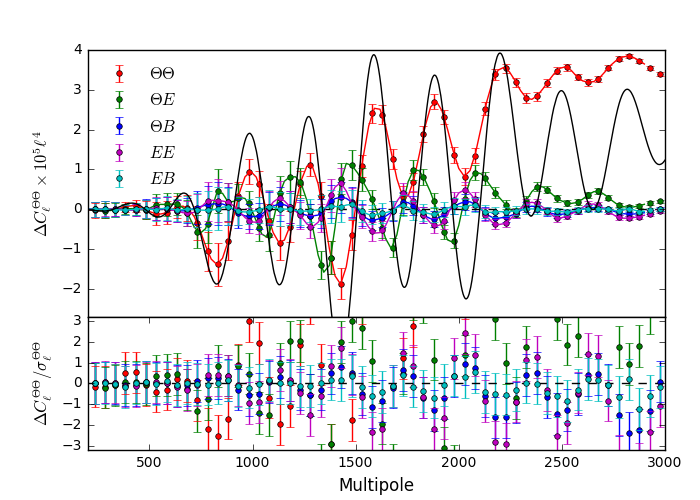} 
\includegraphics[width=8.7cm,height=6cm,clip]{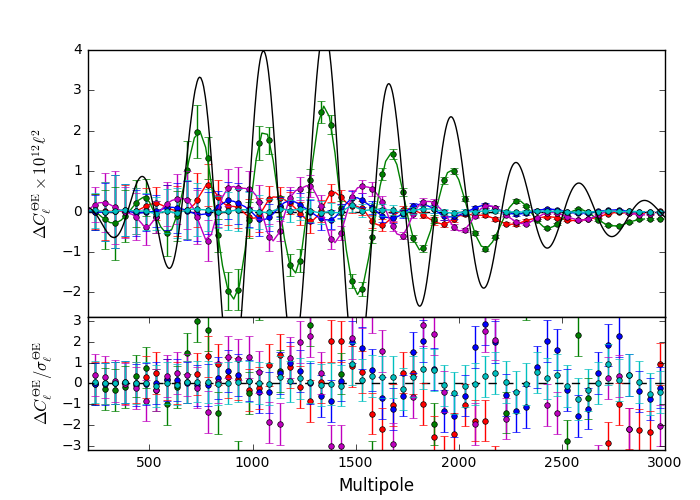} 
\\
\includegraphics[width=8.7cm,height=6cm,clip]{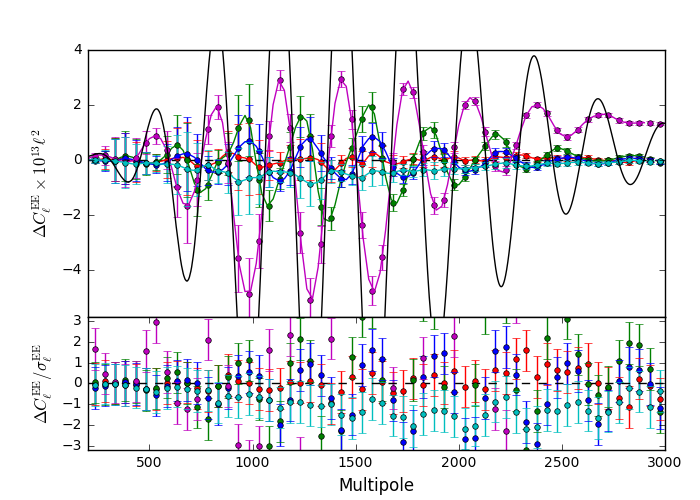} 
\includegraphics[width=8.7cm,height=6cm,clip]{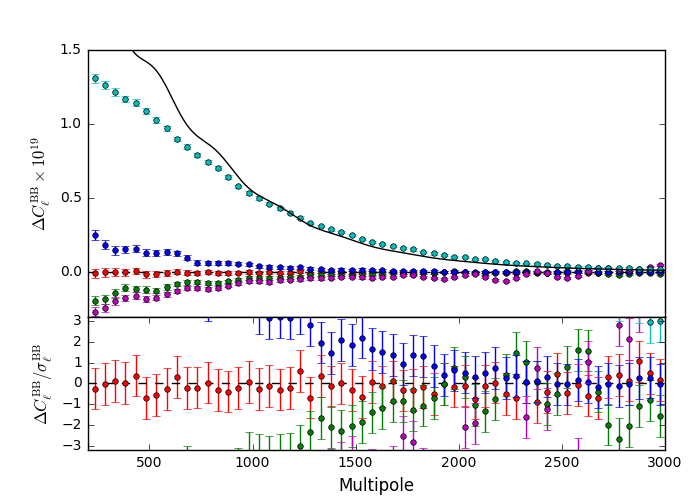} 
\caption{
Delensing bias in $\CTT_\l$, $\CTE_\l$, $\CEE_\l$, and $\CBB_\l$ with reconstructed lensing potentials 
from $\T\T$ (red), $\T E$ (green), $\T B$ (blue), $EE$ (magenta), and $EB$ (cyan) quadratic estimators of \Ref{Hu:2001kj}. 
The solid lines show the difference between the lensed and unlensed CMB power spectra.
We assume $3\mu$K-arcmin and $\sqrt{2}\times 3\mu$K-arcmin white noises in the temperature and 
polarization maps, respectively, and the observed CMB maps are convolved with $1$ arcmin Gaussian beam. 
The error denotes the $1\sigma$ statistical error for one realization of a full sky observation. 
}
\label{fig:tt,te,ee,bb}
\ec
\end{figure*}

In this section, we show the delensing biases in the delensed temperature, $E$, and $B$ mode 
auto power spectra and temperature-$E$-mode cross power spectrum. 
We reconstruct lensing potentials from $\T\T$, $\T E$, $\T B$, $EE$, and $EB$ quadratic estimators of \Ref{Hu:2001kj}. 
The temperature, $E$, and $B$ modes are delensed by the lensing potentials with the remapping method. 

\fig{fig:tt,te,ee,bb} shows the delensing biases using a lensing potential reconstructed from these five estimators. 
The delensing biases are defined as $\Delta C_\l= C_\l^{\rm res,\grad^{\rm inp}}-C_\l^{\rm res}$
where $C_\l^{\rm res}$ and $C_\l^{\rm res,\grad^{\rm inp}}$ are the delensed spectra using 
a reconstructed lensing potential and the input lensing potential with the random Gaussian reconstruction noise, respectively. 
We also show the difference between the lensed and unlensed power spectra. 
In the delensed $\CTT_\l$, $\CTE_\l$, $\CEE_\l$, and $\CBB_\l$, 
the most significant delensing biases come from $\T\T$, $\T E$, $EE$, and $EB$ quadratic estimators, respectively. 

\section{Derivation of the suboptimal polyspectra estimator} \label{app:suboptderiv}

Here we derive \eq{Eq:subopt} by generalizing the discussion of \Ref{Carron:2017} to the case 
with the six point correlation. In the six point correlation, errors in the covariance and trispectrum 
could lead to errors in the estimate of the delensing bias. 
To remove the errors, we consider the following estimator: 
\al{
	\hk'_{i_1\cdots i_6} &= K^{(2,4)}_{i_1\cdots i_6}+K^{(2,2,2)}_{i_1\cdots i_6} 
		+ \sum_{jk} \PD{[K^{(2,4)}_{i_1\cdots i_6}+K^{(2,2,2)}_{i_1\cdots i_6}]}{c_{jk}}\delta c_{jk}
		+ \sum_{jklm} \PD{K^{(2,4)}_{i_1\cdots i_6}}{t_{jklm}}\delta t_{jklm}
	\,. \label{Eq:k:deriv}
}
where we define $c_{ij}=\ave{\ox_i\ox_j}$, $t_{jklm}=\ave{\ox_j\ox_k\ox_l\ox_m}_{\rm c}$, and
\al{
	K^{(2,4)}_{i_1\cdots i_6} &\equiv c_{i_1i_2}t_{i_3i_4i_5i_6} + \text{(14 perms.)} 
	\,, \\
	K^{(2,2,2)}_{i_1\cdots i_6} &\equiv c_{i_1i_2}c_{i_3i_4}c_{i_5i_6} + \text{(14 perms.)} = p^6_3
	\,, \\
	\delta c_{jk} &\equiv \ox_j\ox_k-c_{jk}
	\,, \\
	\delta t_{jklm} &\equiv \ox_j\cdots\ox_m-[\ox_j\ox_kc_{lm}+(\text{5 perms.})]+[c_{jk}c_{lm}+(\text{2 perms.})]-t_{jklm}
	\,.
}
The sum of $K^{(2,4)}$ and $K^{(2,2,2)}$ is the disconnected part of the six point correlation. 
The expectation of the above estimator is equal to the first term, $K^{(2,4)}+K^{(2,2,2)}$. 
If there are errors in the covariance and trispectrum, $\delta c$ and $\delta t$, the 
above estimator does not contain biases from the linear order of $\delta c$ and $\delta t$. 

To simplify the above estimator, we use 
\al{
	\sum_{jk}\PD{K^{(2,4)}_{i_1\cdots i_6}}{c_{jk}}\delta c_{jk}
		&= (\ox_{i_1}\ox_{i_2}-c_{i_1i_2})t_{i_3i_4i_5i_6} + \text{(14 perms.)}
	\,, \\
	\sum_{jk}\PD{K^{(2,2,2)}_{i_1\cdots i_6}}{c_{jk}}\delta c_{jk}
		&= [(\ox_{i_1}\ox_{i_2}-c_{i_1i_2})c_{i_3i_4}c_{i_5i_6}
		+ (\ox_{i_3}\ox_{i_4}-c_{i_3i_4})c_{i_1i_2}c_{i_5i_6}
		+ (\ox_{i_5}\ox_{i_6}-c_{i_5i_6})c_{i_1i_2}c_{i_3i_4}]
		+ \text{(14 perms.)} 
	\,, \\
	\sum_{jklm} \PD{K^{(2,4)}_{i_1\cdots i_6}}{t_{jklm}}\delta t_{jklm}
		&= c_{i_1i_2}\{\ox_{i_3}\cdots\ox_{i_6}-[\ox_{i_3}\ox_{i_4}c_{i_5i_6}+(\text{5 perms.})]+[c_{i_3i_4}c_{i_5i_6}+(\text{2 perms.})]-t_{i_3\cdots i_6}\}
		+ \text{(14 perms.)}
	\notag \\
		&= p_1^6 - 2p_2^6 + 3p_3^6 - K^{(2,4)}_{i_1\cdots i_6}
	\,.
}
Substituting the above equations into \eq{Eq:k:deriv}, we find
\al{
	\hk'_{i_1\cdots i_6} = [(\ox_{i_1}\ox_{i_2}-c_{i_1i_2})\ave{\ox_{i_3}\ox_{i_4}\ox_{i_5}\ox_{i_6}}+(\text{14 perms.})]
		+ p_1^6 - 2p_2^6 + 4p_3^6
	\,.
}
This estimator coincides with \eq{Eq:subopt}. 

\section{Derivation of the RD estimator for delensing bias} \label{app:bias}

In this section, we derive the RD estimator in general cases of the CMB internal delensing. 
We consider delensed CMB anisotropies $\hX$ using a reconstructed lensing potential $\estg^{YZ}$ 
obtained from observed CMB anisotropies, $\hY$ and $\hZ$. 

\subsection{Four point correlation} \label{app:4pt-bias}

We first discuss the RD estimator for the disconnected part of the four point correlation $N^{(4),d}$. 
In the CMB internal delensing, the delensing bias from the disconnected part of the four point correlation is given by
\al{
	N^{(4),d} = -2\ave{(\hX\o\estg^{YZ})\hW^*} \,.
}
From \eq{Eq:opt:kd}, the optimal trispectrum estimator for the disconnected part of $-2(\hX\o\estg^{YZ})\hW^*$ 
is given as $p^4_1-p^4_2$ where $p^4_m$ is defined in \eq{Eq:pnm}. The quantity, $p^4_1$, becomes
\al{
	p^4_1 = -2\ave{2(\hX_1\o\estg_{01})\hW^* + 2(\hX\o\estg_{01})\hW_1^* + (\hX_1\o\estg)\hW_1^*}_{\bm 1}
	\,. \label{Eq:p41}
}
Here we define a symmetric estimator, 
\al{
	2\estg_{ij} = \estg^{Y_iZ_j} + \estg^{Y_jZ_i} 
	\,, \label{Eq:phi-sym}
}
with $i=0$ meaning the real data, and ignore the mean-field bias because its contribution is 
absorbed into the mean-field corrected estimator $\estg=\estg^{YZ}-\mg$ after combining $p_m$. 
We then replace $\estg^{YZ}$ with $\estg$. 
The last term is the correction for the mean-field bias in $(\hX\o\estg)\hW^*$. 
The term, $p^4_2$, becomes
\al{
	p^4_2 = -4\ave{(\hX_1\o\estg_{12})\hW_2^*}_{\bm 1,2} \,. \label{Eq:p42}
}
Using Eqs.~\eqref{Eq:p41} and \eqref{Eq:p42}, the optimal estimator for $N^{(4),d}$ is written as
\al{
	\hN^{(4),d} &= p^4_1-p^4_2 
		\notag \\ 
		&= -2\ave{[\hX_1\o\estg+\hX_{012}\o(\estg^{Y_{0,-2}Z_1}+\estg^{Y_1Z_{0,-2}})]\hW^*_{01}}_{\bm 1,2}
	\,, \label{Eq:hN4d:g}
}
where we define $X_{i,\pm j\dots}=X_i\pm X_j+\dots$. 

The RD estimator of \eq{Eq:hN4d} is derived by substituting $X=Y=E$, $Z=W=B$ into \eq{Eq:hN4d:g}. 
For the temperature delensing with the temperature quadratic estimator, the RD estimator is obtained 
by substituting $X=Y=Z=W=\T$ into \eq{Eq:hN4d:g} and the result is
\al{
	\hN^{(4),d} = -2\ave{[\hT_1\o\estg+2\hT_{012}\o\estg^{\T_{0,-2}\T_1}]\hT^*_{01}}_{\bm 1,2} \,,
}
where $\estg^{\T\T}$ is the temperature quadratic estimator \cite{Hu:2001}. 

\subsection{Six point correlation} \label{app:6pt-opt}

Next we consider the estimator for the six point correlation. For clarity, we omit the operation, $\ave{\cdots}_{\bm i}$. 
In the CMB internal delensing using the template method, the six point correlation is written as $|\hX\o\estg^{YZ}|^2$.
We first derive the estimator based on \eq{Eq:opt:k} and the suboptimal estimator of \eq{Eq:subopt} is 
derived in the next section. 

According to the optimal estimator of \eq{Eq:opt:k}, the RD estimator for the disconnected part 
of the six point correlation is given as $p^6_1-p^6_2+p^6_3$. 
The quantity, $p^6_1$, has $15$ terms and becomes
\al{
	p^6_1 &= 2\Re{\hX_1\o\estg^{Y_1Z}(\hX\o\estg)^*} + 2\Re{\hX_1\o\estg(\hX\o\estg^{Y_1Z})^*}
		+ |\hX_1\o\estg|^2 + |\hX\o\estg^{Y_1Z}|^2 + |\hX\o\estg^{YZ_1}|^2
		\notag \\
		&\quad+ 2\Re{\hX_1\o\estg^{YZ_1}(\hX\o\estg)^*} + 2\Re{\hX_1\o\estg(\hX\o\estg^{YZ_1})^*}
		+ 2\Re{\hX\o\estg^{Y_1Z}(\hX\o\estg^{YZ_1})^*}
	\,.
}
Using the symmetric estimator of \eq{Eq:phi-sym}, the above quantity is reduced to
\al{
	p^6_1 &= 2\Re{\hX_1\o(\estg^{Y_1Z}+\estg^{YZ_1})(\hX\o\estg)^*} + 2\Re{\hX_1\o\estg(\hX\o(\estg^{Y_1Z}+\estg^{YZ_1}))^*}
			+ |\hX_1\o\estg|^2 + |\hX\o(\estg^{Y_1Z}+\estg^{YZ_1})|^2
		\notag \\
		&= 4\Re{\hX_1\o\estg_{01}(\hX\o\estg)^*} + 4\Re{\hX_1\o\estg(\hX\o\estg_{01})^*} + |\hX_1\o\estg|^2 + 4|\hX\o\estg_{01}|^2
	\,. \label{Eq:N6:p61}
}
The quantity, $p^6_2$, consists of $45$ terms. The terms which do not have $\mg$ are given by
\al{
	p^6_2 &= \hX\o\estg^{YZ_1}(\hX_2\o\estg^{Y_1Z_2})^* + \hX\o\estg^{YZ_1}(\hX_2\o\estg^{Y_2Z_1})^*
		\notag \\
		&\quad+ \hX\o\estg^{Y_1Z}(\hX_2\o\estg^{Y_1Z_2})^* + \hX\o\estg^{Y_1Z}(\hX_2\o\estg^{Y_2Z_1})^*
		\notag \\
		&\quad+ \hX\o\estg^{Y_1Z_2}(\hX\o\estg^{Y_2Z_1})^* + |\hX\o\estg^{Y_1Z_2}|^2
		\notag \\
		&\quad+ \hX\o\estg^{Y_1Z_2}(\hX_2\o\estg^{YZ_1})^* + \hX\o\estg^{Y_1Z_2}(\hX_1\o\estg^{YZ_2})^*
		\notag \\
		&\quad+ \hX\o\estg^{Y_1Z_2}(\hX_1\o\estg^{Y_2Z})^* + \hX\o\estg^{Y_1Z_2}(\hX_2\o\estg^{Y_1Z})^*
		\notag \\
		&\quad+ \hX_1\o\estg(\hX_2\o\estg^{Y_2Z_1})^* + \hX_1\o\estg(\hX_2\o\estg^{Y_1Z_2})^*
		\notag \\
		&\quad+ \hX_1\o\estg^{YZ_2}(\hX\o\estg^{Y_2Z_1})^* + \hX_1\o\estg^{YZ_2}(\hX\o\estg^{Y_1Z_2})
		\notag \\
		&\quad+ \hX_1\o\estg^{YZ_1}(\hX_2\o\estg^{YZ_2})^* + \hX_1\o\estg^{YZ_2}(\hX_2\o\estg^{YZ_1})^* + |\hX_1\o\estg^{YZ_2}|^2
		\notag \\
		&\quad+ \hX_1\o\estg^{YZ_1}(\hX_2\o\estg^{Y_2Z})^* + \hX_1\o\estg^{YZ_2}(\hX_2\o\estg^{Y_1Z})^* + \hX_1\o\estg^{YZ_2}(\hX_1\o\estg^{Y_2Z})^*
		\notag \\
		&\quad+ \hX_1\o\estg^{Y_2Z}(\hX\o\estg^{Y_1Z_2})^* + \hX_1\o\estg^{Y_2Z}(\hX\o\estg^{Y_2Z_1})^*
		\notag \\
		&\quad+ \hX_1\o\estg^{Y_2Z}(\hX_1\o\estg^{YZ_2})^* + \hX_1\o\estg^{Y_2Z}(\hX_2\o\estg^{YZ_1})^* + \hX_1\o\estg^{Y_1Z}(\hX_2\o\estg^{YZ_2})^*
		\notag \\
		&\quad+ \hX_1\o\estg^{Y_1Z}(\hX_2\o\estg^{Y_2Z})^* + \hX_1\o\estg^{Y_2Z}(\hX_2\o\estg^{Y_1Z})^* + |\hX_1\o\estg^{Y_2Z}|^2
		\notag \\
		&\quad+ \hX_1\o\estg^{Y_2Z_1}(\hX\o\estg^{YZ_2})^* + \hX_1\o\estg^{Y_2Z_1}(\hX\o\estg^{YZ_1})^*
		\notag \\
		&\quad+ \hX_1\o\estg^{Y_2Z_1}(\hX\o\estg^{Y_2Z})^* + \hX_1\o\estg^{Y_1Z_2}(\hX\o\estg^{Y_2Z})^*
		\notag \\
		&\quad+ \hX_1\o\estg^{Y_2Z_1}(\hX_2\o\estg)^* + \hX_1\o\estg^{Y_1Z_2}(\hX_2\o\estg)^*
	\,.
}
To reduce the above terms, we first rewrite the above equation as
\al{
	p^6_2 &= 2\Re{\hX\o\estg^{YZ_1}(\hX_2\o\estg^{Y_1Z_2})^*} + 2\Re{\hX\o\estg^{YZ_1}(\hX_2\o\estg^{Y_2Z_1})^*}
		\\ \notag
		&\quad+ 2\Re{\hX\o\estg^{Y_1Z}(\hX_2\o\estg^{Y_1Z_2})^*} + 2\Re{\hX\o\estg^{Y_1Z}(\hX_2\o\estg^{Y_2Z_1})^*}
		\\ \notag
		&\quad+ \hX\o\estg^{Y_1Z_2}(\hX\o\estg^{Y_2Z_1})^* + |\hX\o\estg^{Y_1Z_2}|^2 + 2\Re{\hX_1\o\estg^{YZ_2}(\hX\o\estg^{Y_1Z_2})^*}
		\\ \notag
		&\quad+ 2\Re{\hX\o\estg^{Y_1Z_2}(\hX_2\o\estg^{YZ_1})^*}
		\\ \notag
		&\quad+ 2\Re{\hX\o\estg^{Y_1Z_2}(\hX_1\o\estg^{Y_2Z})^*} + 2\Re{\hX\o\estg^{Y_1Z_2}(\hX_2\o\estg^{Y_1Z})^*}
		\\ \notag
		&\quad+ 2\Re{\hX_1\o\estg(\hX_2\o\estg^{Y_2Z_1})^*} + 2\Re{\hX_1\o\estg(\hX_2\o\estg^{Y_1Z_2})^*}
		\\ \notag
		&\quad+ \hX_1\o\estg^{YZ_1}(\hX_2\o\estg^{YZ_2})^* + \hX_1\o\estg^{YZ_2}(\hX_2\o\estg^{YZ_1})^* + |\hX_1\o\estg^{YZ_2}|^2
		\\ \notag
		&\quad+ 2\Re{\hX_1\o\estg^{YZ_1}(\hX_2\o\estg^{Y_2Z})^*} + 2\Re{\hX_1\o\estg^{YZ_2}(\hX_2\o\estg^{Y_1Z})^*} 
			+ 2\Re{\hX_1\o\estg^{YZ_2}(\hX_1\o\estg^{Y_2Z})^*}
		\\ \notag
		&\quad+ \hX_1\o\estg^{Y_1Z}(\hX_2\o\estg^{Y_2Z})^* + \hX_1\o\estg^{Y_2Z}(\hX_2\o\estg^{Y_1Z})^* + |\hX_1\o\estg^{Y_2Z}|^2
	\,.
}
Using the symmetric estimator of \eq{Eq:phi-sym}, we obtain
\al{
	p^6_2 &= 2\Re{\hX\o(\estg^{YZ_1}+\estg^{Y_1Z})(\hX_2\o(\estg^{Y_1Z_2}+\estg^{Y_2Z_1}))^*}
		+ 2\Re{\hX_1\o(\estg^{YZ_2}+\estg^{Y_2Z})(\hX\o(\estg^{Y_1Z_2}+\estg^{Y_2Z_1}))^*}
		\notag \\
		&\quad + 2\Re{\hX_1\o\estg(\hX_2\o(\estg^{Y_2Z_1}+\estg^{Y_1Z_2}))^*}
		\notag \\
		&\quad + \Re{\hX_1\o(\estg^{Y_1Z}+\estg^{YZ_1})(\hX_2\o(\estg^{Y_2Z}+\estg^{YZ_2}))^*}
		+ \Re{\hX_1\o(\estg^{YZ_2}+\estg^{Y_2Z})(\hX_2\o(\estg^{Y_1Z}+\estg^{YZ_1}))^*}
		\notag \\
		&\quad+ \hX\o\estg^{Y_1Z_2}(\hX\o(\estg^{Y_2Z_1}+\estg^{Y_1Z_2}))^* + |\hX_1\o(\estg^{Y_2Z}+\estg^{YZ_2})|^2
	\notag \\
	&= 8\Re{\hX\o\estg_{01}(\hX_2\o\estg_{12})^*} + 8\Re{\hX_1\o\estg_{02}(\hX\o\estg_{12})^*} 
		\notag \\
		&\quad+ 4\Re{\hX_1\o\estg(\hX_2\o\estg_{12})^*}
		\notag \\
		&\quad+ 4\Re{\hX_1\o\estg_{01}\o(\hX_2\o\estg_{02})^*} + 4\Re{\hX_1\o\estg_{02}(\hX_2\o\estg_{01})^*}
		\notag \\
		&\quad+ 2|\hX\o\estg_{12}|^2 + 4|\hX_1\o\estg_{02}|^2
	\,, \label{Eq:N6:p62}
}
where we use $\hX\o\estg^{Y_1Z_2}(\hX\o(\estg^{Y_2Z_1}+\estg^{Y_1Z_2}))^*=2|\hX\o\estg_{12}|^2$. 
The quantity, $p^6_3$, is given by
\al{
	p^6_3 &= 2\hX_1\o\estg^{Y_1Z_2}(\hX_3\o\estg_{23})^* + 2\hX_1\o\estg^{Y_2Z_1}(\hX_3\o\estg_{23})^*
		\notag \\
		&\quad+ 2\hX_1\o\estg^{Y_2Z_3}(\hX_1\o\estg_{23})^* + 2\hX_1\o\estg^{Y_2Z_3}(\hX_3\o\estg_{12})^* 
		\notag \\
		&\quad+ 2\hX_1\o\estg^{Y_2Z_3}(\hX_2\o\estg_{13})^*
	\notag \\
	&= 4\hX_1\o\estg_{13}(\hX_2\o\estg_{23})^* + 4\hX_1\o\estg_{23}(\hX_2\o\estg_{13})^* + 2|\hX_1\o\estg_{23}|^2
	\,. \label{Eq:N6:p63}
}
Combining the above quantities, we obtain
\al{
	\hN &\equiv 4\Re{\hX_1\o\estg_{01}(\hX\o\estg)^*} + 4\Re{\hX_1\o\estg(\hX\o\estg_{01})^*}
		+ |\hX_1\o\estg|^2 + 4|\hX\o\estg_{01}|^2
		\notag \\
		&\quad- 8\Re{\hX\o\estg_{01}(\hX_2\o\estg_{12})^*} - 8\Re{\hX_1\o\estg_{02}(\hX\o\estg_{12})^*} 
		- 4\Re{\hX_1\o\estg(\hX_2\o\estg_{12})^*}
		\notag \\
		&\quad- 4\Re{\hX_1\o\estg_{01}(\hX_2\o\estg_{02})^*} - 4\Re{\hX_1\o\estg_{02}(\hX_2\o\estg_{01})^*} - 2|\hX\o\estg_{12}|^2
		\notag \\
		&\quad- 4|\hX_1\o\estg_{02}|^2 + 4\hX_1\o\estg_{13}(\hX_2\o\estg_{23})^* + 4\hX_1\o\estg_{23}(\hX_2\o\estg_{13})^* + 2|\hX_1\o\estg_{23}|^2
	\,. \label{Eq:p:comb}
}
Using $\hX_{i,\pm j,\dots}=\hX_i\pm\hX_j+\dots$ with $i=0$ denoting the real data, we find
\al{
	4\Re{\hX_1\o\estg_{01}(\hX\o\estg)^*} + 4\Re{\hX_1\o\estg(\hX\o\estg_{01})^*} &= 4\Re{\hX_{012}\o\estg_{01}(\hX_{01}\o\estg)^*}
	\notag \\
	-8\Re{\hX\o\estg_{01}(\hX_2\o\estg_{12})^*} - 8\Re{\hX_1\o\estg_{02}(\hX\o\estg_{12})^*} &= -8\Re{\hX_{012}\o\estg_{13}(\hX_{01}\o\estg_{03})^*}
	\notag \\
	-4\Re{\hX_1\o\estg_{01}(\hX_2\o\estg_{02})^*} - 4\Re{\hX_1\o\estg_{02}(\hX_2\o\estg_{01})^*} &= -4\Re{\hX_{012}\o\estg_{01}(\hX_{12}\o\estg_{02})^*}
	\notag \\
	4\hX_1\o\estg_{13}(\hX_2\o\estg_{23})^* + 4\hX_1\o\estg_{23}(\hX_2\o\estg_{13})^* &= 4\Re{\hX_{012}\o\estg_{13}(\hX_{12}\o\estg_{23})^*}
	\notag \\
	-4\Re{\hX_1\o\estg(\hX_2\o\estg_{12})^*} &= -4\Re{\hX_{012}\o\estg_{13}(\hX_3\o\estg)^*}
	\,. \label{Eq:p2:5}
}
Here we use the fact that the left-hand side of the above fourth equation is real, 
and we also exchange the index ($1\leftrightarrow 3$) in the right-hand side of the last equation.
The sum of the above terms becomes
\al{
	\Re{\hX_{012}\o(2\estg_{01}+2\estg_{13})[\hX_{012}\o(2\estg-2\estg_{02}-4\estg_{03}+2\estg_{23})-2\hX_3\o\estg]^*}
	\,. \label{Eq:p2:cross}
}
The other terms in \eq{Eq:p:comb} except $|\hX_1\o\estg|^2$ are summarized as
\al{
	&- 2|\hX\o\estg_{23}|^2 + 2|\hX_1\o\estg_{23}|^2 + 4|\hX\o\estg_{02}|^2 - 4|\hX_1\o\estg_{02}|^2
		= 2\hX_{012}\o(\estg_{02}+\estg_{23})[\hX_{0,-1}\o(2\estg_{02}-\estg_{23})]^*
	\,. \label{Eq:p2:res}
}
The sum of \eqs{Eq:p2:cross,Eq:p2:res} becomes
\al{
	\Re{\hX_{012}\o(2\estg_{01}+2\estg_{13})\mC{B}_{0123}^*} \,,
}
where we define a realization dependent $B$ mode as
\al{
	\mC{B}_{0123} \equiv \hX_{012}\o(2\estg-2\estg_{02}-4\estg_{03}+2\estg_{23})-2\hX_3\o\estg+\hX_{0,-2}\o(2\estg_{01}-\estg_{13})
}
The estimator for the disconnected part of the six point correlation is recast as
\al{
	\hN = \Re{\hX_{012}\o(2\estg_{01}+2\estg_{13})\mC{B}^*} + |\hX_1\o\estg|^2 \,. 
}

To obtain the delensed power spectrum of \eq{Eq:resBB}, we need to add the term $|\hX_2\o\estg_{11}|^2$ to 
the above estimator because $\hN$ is the estimator for the all disconnected parts of 
the six point correlation and $|\hX_2\o\estg_{11}|^2$ is involved in $\hN$. To recover this term, we add
\al{
	|\hX_2\o\estg|^2 + |\hX\o\estg_{11}|^2 - |\hX_2\o\estg_{11}|^2 \,. \label{Eq:app:abias}
}
This correction is motivated by the following estimator \cite{Carron:2017}
\al{
	\widehat{f} = \ave{f} + \PD{\ave{f}}{x}(x-\ave{f}) \,.
}
where $x$ is any quantity and $f$ is a function of $x$. Subtracting \eq{Eq:app:abias} from $\hN$, we obtain
\al{
	\hN^{(6),d} &\equiv \Re{\hX_{012}\o(2\estg_{01}+2\estg_{13})\mC{B}_{0123}^*}
		+ \hX_{0,-3}\o\estg_{11}(\hX_{03}\o\estg_{11})^*
	\,. \label{Eq:hN6d:general}
}
The RD estimator of \eq{Eq:hN6d} is derived by substituting $X=Y=E$, $Z=B$ into \eq{Eq:hN6d:general}.

\subsection{Six point correlation with suboptimal estimator} \label{app:6pt-sub}

Next we derive the suboptimal estimator for the six point correlation. 
From \eq{Eq:subopt}, the suboptimal estimator is given by
\al{
	\hN^{(6),\rm d,sub} 
		&= p_1^6 - 2p_2^6 + 4p_3^6 
		+ [(\hX\estg^{XY_2}-\hX_1\estg^{Y_1Z_2})\hX_2\estg^{Y_2Z_2} + \text{(14 perms.)}] 
	\,. \label{Eq:N6sub:1}
}
Using \eqs{Eq:N6:p61,Eq:N6:p62,Eq:N6:p63}, we obtain
\al{
	p_1^6 - 2p_2^6 + 4p_3^6 
		&= 4\Re{\hX_1\o\estg_{01}(\hX\o\estg)^*} + 4\Re{\hX_1\o\estg(\hX\o\estg_{01})^*} + 4|\hX\o\estg_{01}|^2 + |\hX_1\o\estg|^2
		\notag \\
		&\quad- 16\Re{\hX\o\estg_{01}(\hX_2\o\estg_{12})^*} - 16\Re{\hX_1\o\estg_{02}(\hX\o\estg_{12})^*} 
		\notag \\
		&\quad- 8\Re{\hX_1\o\estg(\hX_2\o\estg_{12})^*}
		- 8\Re{\hX_1\o\estg_{01}\o(\hX_2\o\estg_{02})^*} -8\Re{\hX_1\o\estg_{02}(\hX_2\o\estg_{01})^*}
		\notag \\
		&\quad- 4|\hX\o\estg_{12}|^2 - 8|\hX_1\o\estg_{02}|^2
		\notag \\
		&\quad+ 4\left[4\hX_{12}\o\estg_{13}(\hX_{12}\o\estg_{23})^* + 2|\hX_1\o\estg_{23}|^2\right]
	\,. \label{Eq:N6sub:p}
}
The last term of \eq{Eq:N6sub:1} is given by
\al{
	&4\Re{\hX\o\estg_{01}(\hX_1\o\estg_{11})^*} + 4\Re{\hX\o\estg_{11}(\hX_1\o\estg_{01})^*} + 4|\hX_1\o\estg_{01}|^2
		\notag \\
	&\quad- 4\Re{\hX_1\o\estg_{12}(\hX_2\o\estg_{22})^*} - 4\Re{\hX_1\o\estg_{22}(\hX_2\o\estg_{12})^*} - 4|\hX_1\o\estg_{12}|^2
		+ |\hX\o\estg_{11}|^2-|\hX_1\o\estg_{22}|^2
	\,. \label{Eq:N6sub:q}
}
Combining \eqs{Eq:N6sub:p,Eq:N6sub:q}, we obtain
\al{
	\hN^{(6),\rm d,sub} 
		&= 4\Re{\hX_1\o\estg_{01}(\hX\o\estg)^*} + 4\Re{\hX_1\o\estg(\hX\o\estg_{01})^*} + 4|\hX\o\estg_{01}|^2
		\notag \\
		&\quad- 16\Re{\hX\o\estg_{01}(\hX_2\o\estg_{12})^*} - 16\Re{\hX_1\o\estg_{02}(\hX\o\estg_{12})^*} 
		\notag \\
		&\quad- 8\Re{\hX_1\o\estg(\hX_2\o\estg_{12})^*}
		- 8\Re{\hX_1\o\estg_{01}\o(\hX_2\o\estg_{02})^*} -8\Re{\hX_1\o\estg_{02}(\hX_2\o\estg_{01})^*}
		\notag \\
		&\quad- 4|\hX\o\estg_{12}|^2 - 8|\hX_1\o\estg_{02}|^2
		\notag \\
		&\quad+ 4\left[4\hX_{12}\o\estg_{13}(\hX_{12}\o\estg_{23})^* + 2|\hX_1\o\estg_{23}|^2\right]
		\notag \\
		&\quad+ 4\Re{\hX\o\estg_{01}(\hX_1\o\estg_{11})^*} + 4\Re{\hX\o\estg_{11}(\hX_1\o\estg_{01})^*} + 4|\hX_1\o\estg_{01}|^2
		\notag \\
		&\quad- 4\Re{\hX_1\o\estg_{12}(\hX_2\o\estg_{22})^*} - 4\Re{\hX_1\o\estg_{22}(\hX_2\o\estg_{12})^*} - 4|\hX_1\o\estg_{12}|^2
	\,, 
}
where $|\hX_1\o\estg|^2+|\hX\o\estg_{11}|^2-|\hX_1\o\estg_{22}|^2$ is subtracted from the above equation. 

To simplify the above equation, we use
\al{
	4|\hX\o\estg_{01}|^2 - 4|\hX\o\estg_{12}|^2 - 8|\hX_1\o\estg_{02}|^2 + 8|\hX_1\o\estg_{23}|^2 
		= \Re{\hX_{012}\o(2\estg_{02}+2\estg_{23})(2(\hX_0-2\hX_1)\o(\estg_{02}-\estg_{23}))^*}
	\,.
}
Using \eq{Eq:p2:5} and the above equation (with exchanging the index, $1\leftrightarrow 2$), we obtain
\al{
	\hN^{(6),\rm d,sub} 
		&= \Re{\hX_{012}\o(2\estg_{01}+2\estg_{13})[\hX_{012}\o(2\estg-4\estg_{02}-8\estg_{03}+8\estg_{23})-4\hX_3\o\estg+2(\hX_0-2\hX_2)\o(\estg_{01}-\estg_{13})]^*}
		\notag \\
		&\quad+ 4\Re{\hX\o\estg_{01}(\hX_1\o\estg_{11})^*} + 4\Re{\hX\o\estg_{11}(\hX_1\o\estg_{01})^*} + 4|\hX_1\o\estg_{01}|^2
		\notag \\
		&\quad- 4\Re{\hX_1\o\estg_{12}(\hX_2\o\estg_{22})^*} - 4\Re{\hX_1\o\estg_{22}(\hX_2\o\estg_{12})^*} - 4|\hX_1\o\estg_{12}|^2
	\,, \label{Eq:N6sub:2}
}
The sum of the second and third lines of the above equation is given by
\al{
	4\Re{\hX_{013}\o(\estg_{01}+\estg_{12})[\hX_{013}\o\estg_{11}-\hX_2\o(\estg_{11}+\estg_{22})+\hX_1\o(\estg_{01}-\estg_{12})]^*}
	\,.
}
Substituting the above equation into \eq{Eq:N6sub:2}, we find
\al{
	\hN^{(6),\rm d,sub} = \Re{\hX_{012}\o(2\estg_{01}+2\estg_{13})(\mC{B}_{0123}^{\rm sub})^*}
	\,, \label{Eq:N6sub:3}
}
where we define a realization-dependent $B$ mode as
\al{
	\mC{B}^{\rm sub}_{0123} 
		&= \hX_{012}\o(2\estg-4\estg_{02}-8\estg_{03}+8\estg_{23}+2\estg_{11})
			-2\hX_3\o(2\estg+\estg_{11}+\estg_{33})+2(\hX_{01}-2\hX_2)\o(\estg_{01}-\estg_{13})
	\,.
}
\eq{Eq:hN6d:sub} is derived with $X=Y=E$ and $Z=B$ in \eq{Eq:N6sub:3}. 

\subsection{Six point correlation relevant to the dominant term of the delensing bias} \label{app:6pt-dom}

Here we consider the RD estimator for \eq{Eq:Ndom}. 
To extract the corresponding terms of \eq{Eq:Ndom} in \eq{Eq:subopt}, we first rewrite \eq{Eq:subopt} as
\al{
	[(\ox_{i_1}\ox_{i_2}-\ave{\ox_{i_1}\ox_{i_2}})(\ave{\ox_{i_3}\cdots\ox_{i_6}}_{\rm c}+\ave{\ox_{i_3}\cdots\ox_{i_6}}_{\rm d})
		+ \ave{\ox_{i_1}\ox_{i_2}}\uk_{i_3i_4i_5i_6} ] + \text{(14 perms.)} + p^6_3
	\,. \label{Eq:subopt-recast}
}
Here $\uk_{\i_3i_4i_5i_6}$ is the unnormalized estimator of the trispectrum defined in \eq{Eq:ukdef}. 

We first consider the term containing the connected four point correlation in \eq{Eq:Ndom}. 
The corresponding terms in \eq{Eq:subopt-recast} are given by
\al{
	(\ox_{i_1}\ox_{i_2}-\ave{\ox_{i_1}\ox_{i_2}})\ave{\ox_{i_3}\cdots\ox_{i_6}}_{\rm c} 
		+ \ave{\ox_{i_1}\ox_{i_2}}\uk_{i_3i_4i_5i_6} 
	\,.
}
The permutations of the above quantity are not relevant to \eq{Eq:Ndom}. 
The above equation is recast as
\al{
	&(\ox_{i_1}\ox_{i_2}-\ave{\ox_{i_1}\ox_{i_2}})(\ave{\ox_{i_3}\cdots\ox_{i_6}}-\ave{\ox_{i_3}\cdots\ox_{i_6}}_{\rm d}) 
	+ \ave{\ox_{i_1}\ox_{i_2}}(\ox_{i_3}\cdots\ox_{i_6}-\ave{\ox_{i_3}\ox_{i_4}}\ox_{i_5}\ox_{i_6}-(\text{5 perms.})+\ave{\ox_{i_3}\cdots\ox_{i_6}}_{\rm d})
	\notag \\
	&= (\ox_{i_1}\ox_{i_2}-\ave{\ox_{i_1}\ox_{i_2}})\ave{\ox_{i_3}\cdots\ox_{i_6}} 
	+ \ave{\ox_{i_1}\ox_{i_2}}(\ox_{i_3}\cdots\ox_{i_6}-\ave{\ox_{i_3}\ox_{i_4}}\ox_{i_5}\ox_{i_6}-(\text{5 perms.}))
	\notag \\
	&\qquad + (2\ave{\ox_{i_1}\ox_{i_2}}-\ox_{i_1}\ox_{i_2})\ave{\ox_{i_3}\cdots\ox_{i_6}}_{\rm d})
	\,.
}
From the above equation, the RD estimator for the second term of \eq{Eq:Ndom} is obtained by 
replacing $\ox_{i_1}\to \hE$, $\ox_{i_2}\ox_{i_3}\to \estg^{EB}$, $\ox_{i_4}\to \hE$ and $\ox_{i_5}\ox_{i_6}\to \estg^{EB}$. 
The result is 
\al{
	\hN^{\rm dom,1} &\equiv 
		(\hE\o\estg^{EB_2}-\hE_1\o\estg^{E_1B_2})(\hE_2\o\estg_{22})^*
	\notag \\
		&\quad+ \hE_1\o\estg^{E_1B}(\hE\o\estg)^* - \hE_1\o\estg^{E_1B}(\hE\o\estg_{22}+2\hE_2\o\estg_{02})^*
		- \hE_1\o\estg^{E_1B_2}(2\hE\o\estg_{02}+\hE_2\o\estg)^*
	\notag \\
		&\quad+ 2(2\hE_1\o\estg^{E_1B_2}-\hE\o\estg^{EB_2})(\hE_3\o\estg_{23})^*
	\\
		&= (\hE\o\estg^{EB_2}-\hE_1\o\estg^{E_1B_2})(\hE_2\o\estg_{22}-2\hE_3\o\estg_{23})^*
	\notag \\
		&\quad+ \hE_1\o\estg^{E_1B}[\hE\o(\estg-\estg_{22}) - 2\hE_2\o\estg_{02}]^*
	\notag \\
		&\quad+ \hE_1\o\estg^{E_1B_2}(2\hE_3\o\estg_{23}-2\hE\o\estg_{02}-\hE_2\o\estg)^*
	\,. \label{Eq:hNdom1}
}
To simplify the above equation, we use
\al{
	\text{(the first line of \eq{Eq:hNdom1})}
		&= \hE_{0,-1}\o\estg^{E_{01}B_{02}}[\hE_{23}\o(\estg_{22}-2\estg_{23})]^*
	\notag \\
	\text{(the second line of \eq{Eq:hNdom1})}
		&= \hE_1\o\estg^{E_{01}B_{02}}[\hE\o(\estg-\estg_{22}) - 2\hE_2\o\estg_{02}]^*
	\notag \\
	\text{(the third line of \eq{Eq:hNdom1})}
		&= \hE_1\o\estg^{E_{01}B_{02}}(2\hE_3\o\estg_{23}-2\hE\o\estg_{02}-\hE_2\o\estg)^*
	\,.
}
Note that the ensemble average of the above first and third equations is zero 
while that of the second equation gives the connected part in \eq{Eq:Ndom}. 
With the above equations, \eq{Eq:hNdom1} becomes 
\al{
	\hN^{\rm dom,1}
		&= \hE_{0,-1}\o\estg^{E_{01}B_{02}}[\hE_{23}\o(\estg_{22}-2\estg_{23})]^*
		+ \hE_1\o\estg^{E_{01}B_{02}}[\hE\o(\estg-\estg_{22}-2\estg_{02})-\hE_2\o(\estg+2\estg_{02})+2\hE_3\o\estg_{23}]^*
	\,.
}

Next we consider the full disconnected part of the six point correlation (i.e., the third term) 
involved in \eq{Eq:Ndom}. The corresponding terms in \eq{Eq:subopt-recast} are described by
\al{
	[(\ox_{i_1}\ox_{i_2}-\ave{\ox_{i_1}\ox_{i_2}})\ave{\ox^{\bm 1}_{i_3}\ox^{\bm 2}_{i_4}\ox^{\bm 2}_{i_5}\ox^{\bm 1}_{i_6}}_{\bm 1,2}
	+ \text{($i_1$,$i_2$)$\leftrightarrow$($i_3$,$i_4$)} + \text{($i_1$,$i_2$)$\leftrightarrow$($i_5$,$i_6$)}]
	+ \ave{\ox_{i_1}\ox_{i_2}}\ave{\ox^{\bm 1}_{i_3}\ox^{\bm 2}_{i_4}\ox^{\bm 2}_{i_5}\ox^{\bm 1}_{i_6}}_{\bm 1,2}
	\,.
}
Replacing $\ox_{i_1}\to \hE$, $\ox_{i_2}\ox_{i_3}\to \estg^{EB}$, $\ox_{i_4}\to \hE$ and $\ox_{i_5}\ox_{i_6}\to \estg^{EB}$, 
we obtain the RD estimator for the third term of \eq{Eq:Ndom} as
\al{
	\hN^{\rm dom,2} 
	&\equiv 2\hE\o\estg^{EB_1}(\hE_2\o\estg^{E_2B_1})^* + \hE_1\o\estg^{E_1B}(\hE_2\o\estg^{E_2B})^* 
		- 2\hE_1\o\estg^{E_1B_2}(\hE_3\o\estg^{E_3B_2})^*
	\notag \\
	&= \hE_1\o\estg^{E_{01}B_{02}}(2\hE\o\estg^{EB_2}+\hE_2\o\estg^{E_2B}-2\hE_3\o\estg^{E_3B_2})^*
	\,.
}

Combining $\hN^{\rm dom,1}$ and $\hN^{\rm dom,2}$, the RD estimator of \eq{Eq:Ndom} is given as
\al{
	\hN^{(6),\rm dom} 
		&= 2\hN^{\rm dom,1} + \hN^{\rm dom,2}
	\notag \\
		&= 2\hE_{0,-1}\o\estg^{E_{01}B_{02}}[\hE_{23}\o(\estg_{22}-2\estg_{23})]^*
	\notag \\
		&\quad+\hE_1\o\estg^{E_{01}B_{02}}[2\hE\o(\estg-\estg_{22}-2\estg_{02}+\estg^{EB_2})
		-\hE_2\o(2\estg+4\estg_{02}-\estg^{E_2B})+2\hE_3\o(2\estg_{23}-\estg^{E_3B_2})]^*
	\notag \\
		&= 2\hE_{0,-1}\o\estg^{E_{01}B_{02}}[\hE_{23}\o(\estg_{22}-2\estg_{23})]^*
	\notag \\
		&\quad+\hE_1\o\estg^{E_{01}B_{02}}[2\hE\o(\estg-\estg_{22}-\estg^{E_2B})
		-\hE_2\o(2\estg+2\estg^{EB_2}+\estg^{E_2B})+2\hE_3\o\estg^{E_2B_3}]^*
	\notag \\
		&= 2\hE_{0,-1}\o\estg^{E_{01}B_{02}}[\hE_{23}\o(\estg_{22}-2\estg_{23})]^*
	\notag \\
		&\quad+\hE_1\o\estg^{E_{01}B_{02}}[2\hE\o(\estg-\estg_{22}-\estg^{E_2B})
		-\hE_2\o(\estg^{(2E+E_2)B_{02}}-\estg_{22})+2\hE_3\o\estg^{E_2B_3}]^*
	\,.
}

\twocolumngrid
\bibliographystyle{apsrev}
\bibliography{exp,main}

\end{document}